\documentclass[letter]{emulateapj}

\usepackage{natbib}

\newcommand{\uK}{$\mu$K$_{CMB}$}
\newcommand{\Msun}{M$_{\odot}$}

\begin{document}

\title{Cluster morphologies and model-independent $Y_{SZ}$ 
  estimates from Bolocam Sunyaev-Zel'dovich images}

\author{
J.~Sayers\altaffilmark{1,3}, S. R. Golwala\altaffilmark{1},
S.~Ameglio\altaffilmark{2}, and E.~Pierpaoli\altaffilmark{2}}
\altaffiltext{1}
  {California Institute of Technology, Pasadena, CA 91125}
\altaffiltext{2}
  {University of Southern California, Los Angeles, CA 90089}
\altaffiltext{3}
  {jack@caltech.edu}

\begin{abstract}
  
  We present initial results from our ongoing program
  to image the Sunyaev-Zel'dovich (SZ) effect in 
  galaxy clusters at 143~GHz using Bolocam;
  five clusters and one blank field are described in this
  manuscript.
  The images have a resolution of 58~arcsec and 
  a radius of $\simeq 6-7$~arcmin,
  which is approximately $r_{500} - 2 r_{500}$ for these clusters.
  We effectively high-pass filter our data in order
  to subtract noise sourced by atmospheric fluctuations,
  but we are able to
  to obtain unbiased images of the clusters
  by deconvolving the effects of this filter.
  The beam-smoothed RMS is $\simeq 10$~\uK~in these images;
  with this sensitivity 
  we are able 
  to detect SZ signal to beyond $r_{500}$
  in binned radial profiles.
  We have fit our images to 
  beta and Nagai models, 
  fixing spherical symmetry or allowing for
  ellipticity in the plane of the sky,
  and we find that the
  best-fit parameter values are in general
  consistent with those obtained 
  from other X-ray and SZ data.
  Our data show no clear preference for the 
  Nagai model or the beta model due
  to the limited spatial dynamic range of our
  images.
  However, our data show a definitive preference
  for elliptical models over spherical models,
  quantified by an $F$-ratio of $\simeq 20$
  for the two models.
  The weighted mean ellipticity of the five
  clusters is $\epsilon = 0.27 \pm 0.03$, 
  consistent with results from X-ray data.
  Additionally, we obtain model-independent
  estimates of $Y_{500}$, the integrated
  SZ $y$-parameter over the cluster face to a radius of $r_{500}$,
  with systematics-dominated uncertainties of $\simeq 10$\%.
  Our $Y_{500}$ values, which are free from the biases
  associated with model-derived $Y_{500}$ values,
  scale with cluster mass in
  a way that is consistent with both self-similar predictions
  and expectations of a $\simeq 10$\% intrinsic scatter.

\end{abstract}

\keywords{cosmology: observation --- galaxies: clusters: individual
   (Abell 697, Abell 1835, MS 0015.9+1609, MS 0451.6-0305, MS 1054.4-0321) ---
   methods: data analysis}

\section{Introduction}

  Galaxy clusters are the largest collapsed objects in the
  universe, making them excellent tools for
  studying cosmology and the astrophysics of
  gravitational collapse.
  They are rare excursions in the matter density
  field and the formation history 
  of clusters is closely tied to the 
  composition and evolution of the universe.
  As a consequence, clusters have been used
  extensively to constrain cosmology.
  For example, they provided the first evidence that
  the matter density, $\Omega_m$,
  was insufficient to close the
  universe \citep{bahcall97}.
  Additionally,
  clusters provided the most
  reliable information about the amplitude
  of scalar perturbations, $\sigma_8$,
  prior to the release of WMAP results
  \citep{viana99, pierpaoli01, pierpaoli03,
    borgani01, allen03a, spergel03}.

  Moreover, the cluster-derived measurements
  of $\sigma_8$
  can be combined with
  measurements of the normalization
  of the CMB power spectrum to constrain the
  total neutrino mass via the growth
  of density perturbations 
  \citep{pierpaoli04, allen03b, vikhlinin09b}.
  In addition, galaxy clusters have been
  forming during the same epoch that dark
  energy has evolved to dominate the energy
  density
  the universe.
  As a result, the
  number of galaxy clusters as a function
  of redshift is sensitive to the
  dark energy density, $\Omega_{\lambda}$, 
  and its equation of state, 
  $w$~\citep{haiman01, holder00, holder01}.
  Recently, large surveys of galaxy clusters
  have been completed, or have released initial
  results, that constrain the 
  total neutrino mass and/or the properties of dark energy
  \citep{allen08, mantz10, vikhlinin09b, vanderlinde10}.
  Additionally, clusters are being used to test gravity
  on large scales via studies of their
  internal structure and distribution
  in space \citep{schmidt09, diaferio09, 
    martino09, wu10, moffat10, rapetti10}.
  
  Galaxy clusters also provide 
  excellent laboratories for studying the 
  astrophysics of structure formation.
  In general, galaxy clusters are 
  well-behaved objects, and their properties
  can be predicted to fairly good precision
  using simple gravitational collapse models
  and self-similar scaling relations~\citep{kaiser86}.
  However, non-gravitational effects, 
  such as radiative cooling, star formation,
  turbulence, magnetic field support, and
  cosmic ray pressure produce deviations
  from the simple gravitational models,
  and the data clearly favor models that
  include non-gravitational 
  processes~\citep{kravtsov06, nagai07}.

  A wide
  range of observational techniques
  are used to study galaxy clusters.
  Optical/infrared and radio measurements
  can be used to study the individual
  galaxies within the cluster
  (\emph{e.g.}, \citet{pipino10,lin09,gralla10});
  parameters such as velocity dispersion
  and richness can then be used to
  understand the global properties of the 
  cluster(\emph{e.g.}, \citet{menanteau10,rines10,serra10,szabo10, hao10}).
  Additionally, X-ray observations
  are sensitive to the bremsstrahlung
  emission from the hot gas in the
  intra-cluster medium (ICM), 
  which contains $\simeq 90$\% of
  the baryonic mass of the cluster
  (\emph{e.g.}, \citet{vikhlinin06,gonzalez07,vikhlinin09b,arnaud09,zhang10}).
  Galaxy clusters are also efficient gravitational
  lenses, and detailed observations of the 
  background galaxies can be used to 
  determine the matter distribution within
  the cluster, $\simeq 90$\% of which is
  in the form of dark matter~\citep{clowe06, allen08}.

  The ICM can also be studied using the 
  Sunyaev-Zel'dovich (SZ) effect~\citep{sunyaev72},
  where the background CMB photons inverse
  Compton scatter off of the electrons
  in the ICM.
  Studies of galaxy clusters using the SZ effect
  are quickly maturing.
  The South Pole Telescope (SPT) and 
  the Atacama Cosmology Telescope (ACT) 
  are conducting large untargeted
  surveys and have already published catalogues
  with dozens of clusters~\citep{vanderlinde10, 
  menanteau10_2}; 
  they expect to detect hundreds of clusters when
  the surveys are complete.
  Due to the redshift independence of the SZ
  surface brightness, the clusters discovered
  in these surveys are, on average, at significantly
  higher redshifts than those discovered with X-ray
  or optical surveys.
  Consequently, SZ-selected cluster catalogues
  are expected to play a leading role
  in further constraining cosmological
  parameters such as $\Omega_{\Lambda}$ and 
  $w$ (\emph{e.g.}, \citet{carlstrom02}).

  These large-scale SZ surveys are operating
  based on the prediction that the total
  integrated SZ signal over the cluster, $Y$,
  which is proportional to the total thermal
  energy of the ICM, scales in a robust way
  with total cluster mass (\emph{e.g.},
  \citet{kravtsov06}).
  In practice, the integral does not generally
  extend to the edge of the cluster, 
  which is effectively at $\gtrsim 5 r_{500}$,
  and can be
  be performed in one of 
  two ways:
  over a spherical volume using a deprojected
  SZ profile or over a cylindrical volume using
  a projected SZ profile.
  Most groups
  have calculated the integrated SZ signal
  by performing the cylindrical integration
  over the cluster face within a well
  defined aperture (generally $r_{2500}$
  or $r_{500}$)~\citep{morandi07,bonamente08, mroczkowski09,
    marrone09,plagge10,huang10,culverhouse10, andersson10},
  although a couple groups have also performed the
  spherical integral~\citep{mroczkowski09,andersson10}.
  With the exception of \citet[hereafter P10]{plagge10},
  these groups have exclusively used parametric profiles
  to describe the SZ signal in order to determine $Y$.
  The initial scaling results from these $Y$
  measurements are roughly consistent with the
  expectation that it is a low-scatter ($\simeq 10$\%)
  proxy for the total cluster mass.

  Additionally, 
  several groups have made detailed SZ observations
  of previously known galaxy clusters,
  and these data have been used to constrain the
  properties of the ICM beyond 
  $r_{500}$~\citep{halverson09, nord09, basu10, 
    mroczkowski09, plagge10}.
  For example, the SPT has measured radial profiles
  in 15 clusters to a significant fraction of the 
  virial radius, and they find best-fit model
  parameters that are consistent with 
  previous studies using X-ray data~P10.
  APEX-SZ has published detailed studies of 
  three clusters, using a joint SZ/X-ray
  analysis to constrain mass-weighted temperature
  profiles beyond $r_{500}$ in two of these
  clusters~\citep{halverson09, nord09, basu10}.
  Note that, in addition to providing complementary information
  about the ICM in these previously studied
  clusters, these SZ data will help constrain
  and improve systematic uncertainties in 
  blind SZ surveys
  (\emph{e.g.}, constraining the typical
  SZ profiles and resulting detection biases
  due to the cooling properties of the 
  central gas \citep{pipino01}).

  High resolution ($\simeq 10$~arcsec) SZ measurements
  are also now being made using MUSTANG~\citep{mason10, korngut10} 
  and SZA/CARMA~\citep{carlstrom02}.
  These observations will provide insights on the
  internal structure of clusters that will be
  complementary to the information obtained 
  via X-ray observations.
  Among other applications,
  these data can be used, in combination
  with optical observations,
  to infer the merging history and evolution
  of the cluster \citep{croston08}.

  In this work, we present Bolocam SZ observations
  of five massive galaxy clusters.
  We are able to measure SZ signal in our 
  two-dimensional images to $\simeq r_{500}$ and
  to well beyond $r_{500}$ in azimuthally averaged
  radial profiles.
  Using these data, we are able to constrain
  the broad morphologies of these clusters
  and compute observables such as 
  $Y_{SZ}$ in a model-independent way.
  A companion paper to this manuscript,
  \citet{ameglio10},
  presents joint X-ray/SZ deprojections
  of density and temperature profiles
  for these same five clusters using
  Bolocam and \emph{Chandra} data
  \citep{ameglio07}.

\section{Cluster observations with Bolocam}
  \label{sec:observations}

\begin{figure}
  \plotone{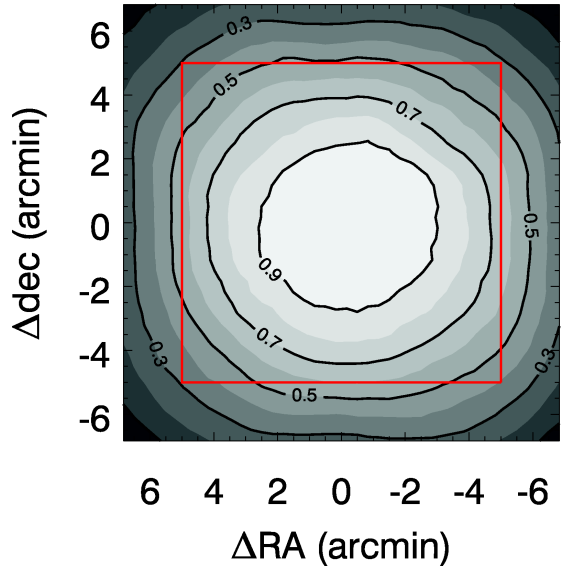}
  \caption{Integration time per pixel, relative
    to the maximum integration time,
    for MS 0451.6-0305.
    Our model fits include all of the
    data within a circular region with
    a minimum integration time of 
    25\% of the peak integration time,
    which corresponds to $6-7$ arcmin
    in radius.
    The red box, with 10~arcmin sides,
    denotes the region used for
    deconvolution of the processing
    transfer function.
    The minimum relative integration
    time within this region is also
    $>25$\%.}
  \label{fig:coverage}
\end{figure}

  Bolocam is a mm-wave imaging camera that operates
  from the Caltech Submillimeter Observatory (CSO)
  with 144 bolometric detectors covering a
  circular 8~arcmin field of view (FOV)~\citep{glenn98, haig04}.
  For the observations described in this manuscript,
  Bolocam was configured to observe at 143~GHz.
  To image the clusters, we scanned the telescope
  in a Lissajous pattern~\citep{kovacs06}, where the telescope
  is driven in two orthogonal directions using
  sine waves with incommensurate periods.
  We used an amplitude of 4~arcmin for the sinusoids,
  which were oriented along the RA and dec directions
  with periods of 6.28 and 8.89~sec.
  These parameters were chosen to keep the FOV
  on the cluster center 100\% of the time while
  scanning as fast as possible at the CSO ($\simeq 4$ arcmin/sec)
  to modulate the cluster signal above the 
  low-frequency atmospheric noise.
  An example of the resulting
  integration time per pixel is given in Figure~\ref{fig:coverage}.
  The data were collected via 10-minute-long observations,
  with the periods of the two orthogonal sinusoids
  exchanged between observations.
  Typically, we complete $\simeq 100$ observations
  per cluster, which corresponds to approximately
  50~ksec and yields a beam-smoothed RMS of 
  $\simeq 10$~\uK~(see Table~\ref{tab:clusters}).

\begin{deluxetable*}{cccccccc} 
  \tablewidth{0pt}
  \tablecaption{Cluster properties}
  \tablehead{\colhead{target} & \colhead{RA} &
    \colhead{dec} & \colhead{redshift} &
    \colhead{Bolocam time (ksec)} & \colhead{RMS (\uK)} &
    \colhead{$r_{500}$ (Mpc)} & 
    \colhead{$M_{gas,500}$ (\Msun)}}
  \startdata
  Abell 697 & 08:42:58 & +36:21:56 & 0.28 & 52 & \phn8.9 & $1.65 \pm 0.09$ & $19.6 \pm 2.7 \times 10^{13}$ \\
  Abell 1835 & 14:01:02 & +02:52:42 & 0.25 & 50 & \phn8.7 & $1.49 \pm 0.06$ & $14.1 \pm 1.2\times 10^{13}$ \\
  MS 0015.9+1609 & 00:18:34 & +16:26:13 & 0.54 & 38 & 10.2 & $1.28 \pm 0.08$ & $17.5 \pm 1.9 \times 10^{13}$ \\
  MS 0451.6-0305 & 04:54:11 & -03:00:53 & 0.55 & 53 & \phn7.7 & $1.45 \pm 0.12$ & $15.6 \pm 2.2 \times 10^{13}$ \\
  MS 1054.4-0321 & 10:56:59 & -03:37:34 & 0.83 & 66 & \phn6.7 & $1.07 \pm 0.13$ & $11.5 \pm 2.4 \times 10^{13}$ \\
  SDS1 & 02:18:00 & -05:00:00 & - & 37 & \phn9.1 & - & -
  \enddata
  \tablecomments{A list of the clusters presented in this 
    manuscript.
    From left to right the columns give the RA and dec of 
    the cluster in J2000 coordinates, the redshift of the
    cluster, the amount of Bolocam integration time,
    the median RMS per beam-smoothed pixel in the Bolocam map,
    the radius of the cluster, and the mass of the cluster.
    The values for $r_{500}$ and $M_{gas,500}$ were taken
    from \citet{mantz10b} (Abell 697 and Abell 1835) and
    \citet{ettori09} (MS 0015.9+1609, MS 0451.6-0305, 
    and MS 1054.4-0321)}
  \label{tab:clusters}
\end{deluxetable*}

  In this paper, we present the results from five clusters
  and one blank field, which are described below.
  Due to the size of our resulting maps ($r \simeq 6-7$~arcmin),
  we have chosen to focus primarily on high-redshift
  clusters, which have virial radii within the extent of our maps.
  All of the clusters presented in this manuscript are 
  beyond $z = 0.25$, and three of the five are
  beyond $z = 0.50$.
  For reference, $r_{500}$ lies within the extent of our map for a typical
  $10^{15}$~\Msun~cluster at $z = 0.25$, 
  and at $z=0.50$, the virial radius lies within
  our map.
  Note that, throughout this work, we assume a $\Lambda$CDM
  cosmology with $\Omega_m = 0.3$, $\Omega_{\Lambda} = 0.7$,
  and $H_0 = 70$~km/s/Mpc.

  \begin{description}
    \item[Abell 697]{
      Abell 697 is a cluster undergoing a complex
      merger event along the line of sight~\citep{girardi06}.}
    \item[Abell 1835]{
      Abell 1835 is a relaxed cluster with a strong cooling flow
      \citep{peterson01, schmidt01}.}
    \item[MS 0015.9+1609]{
      MS 0015.9+1609 
      is a triaxial cluster that is elongated along the
      line of sight and has an anomalously high gas
      mass fraction of 27\% \citep{piffaretti03}.}
    \item[MS 0451.6-0305]{
      MS 0451.6-0305 is a cluster that is not quite in gravitational
      equilibrium, with a slightly elongated X-ray profile
      \citep{donahue03}.}
    \item[MS 1054.4-0321]{
      MS 1054.4-0321 is 
      a cluster undergoing a merger, as
      evidenced by the presence of two distinct sub-clumps 
      in the X-ray image \citep{jeltema01, jee05}.}
    \item[SDS1]{
      Our SDS1 map is centered
      in the middle
      of the Subaru/XMM Deep Survey (SXDS) field.
      The deep \emph{XMM-Newton} survey of this field reveals only
      3 clusters within our map, all near the edge,
      and the largest of which has a virial
      mass of $M_{200} = 0.8 \times 10^{13}$~\Msun~\citep{finoguenov10}.
      Therefore, SDS1 is approximately free of signal
      from the SZ effect.}
  \end{description}

\section{Data reduction}
  \label{sec:reduction}

    In general, our data reduction followed the procedure
    described in \citet[hereafter S09]{sayers09}, 
    with some
    minor modifications.
    We briefly describe the techniques below, along with
    the changes relative to S09.

  \subsection{Calibration}

    Bright quasars located near the clusters were
    observed for $10$~minutes once every $\simeq 90$
    minutes in order to determine the offset
    of our focal plane relative to the telescope
    pointing coordinates.
    These observations were used to construct a model
    of the pointing offset as a function of local
    coordinates (az,el), with a single model for
    each cluster.
    The uncertainty in the pointing models is
    $\lesssim 5$~arcsec.
    This pointing uncertainty is quasi-negligible
    for Bolocam's 58~arcsec FWHM beams, especially
    for extended objects such as clusters.
    We made two 20-minute-long observations
    each night of Uranus, Neptune,
    or a source in \cite{sandell94} for flux
    calibration.
    Using the quiescent detector resistance as 
    a proxy for detector responsivity and atmospheric
    transmission,
    we then fit a single
    flux-calibration curve to the entire data
    set.
    We estimate the uncertainty in our flux
    calibration to be 4.3\%, with the 
    following breakdown:
    1.7\% from the
    Rudy temperature model of Mars scaled to
    measured WMAP values~\citep{halverson09,
      wright76, griffin86, rudy87, muhleman91, hill09},
    1.5\% in the Uranus/Neptune model referenced
    to Mars~\citep{griffin93},
    1.4\% due to variations in atmospheric opacity~(S09),
    3.1\% due to uncertainties in the solid angle
    of our point-spread function (PSF)~(S09),
    and 1.5\% due to measurement uncertainties~(S09).

  \subsection{Atmospheric noise subtraction}

    The raw Bolocam timestreams are dominated
    by noise sourced by fluctuations in the 
    water vapor in the atmosphere,
    which have a power spectrum that rises sharply
    at low frequencies.
    In order to optimally subtract the atmospheric
    noise, we have used a slightly modified version
    of the average subtraction algorithm
    described in \citet[hereafter S10]{sayers10}.
    We have modified the S10 algorithm because
    these cluster data contain 
    additional atmospheric noise caused
    by the Lissajous scan pattern.
    Since we are scanning the telescope parallel to
    RA and dec, the airmass 
    we are looking
    through is constantly changing.
    As a result, our data contain a large amount
    of atmospheric signal in narrow bands centered	
    on the two fundamental scan frequencies.

    Following the algorithm in S10,
    we first create a template of the atmosphere by 
    averaging the signal from all of our detectors
    at each time sample (\emph{i.e.}, the 
    average signal over the FOV). 
    In S10, this template is subtracted from
    each detector's timestream after weighting
    it by the relative gain of that detector,
    which is determined from the correlation
    coefficient between 
    the timestream and the template.
    We use a single correlation coefficient 
    for each detector for each 10-minute-long
    observation.
    However, a significant fraction of the
    atmospheric noise at the fundamental scan
    frequencies remains in the data after application
    of the S10 algorithm, indicating that
    we have slightly misestimated the
    correlation coefficients.
    Therefore, we modified the S10 algorithm to
    compute the correlation coefficients for the template
    based only on the data within a narrow band
    centered on the two fundamental scan frequencies.
    The atmospheric noise power in these narrow
    frequency bands is roughly an order of magnitude above
    the broadband atmospheric noise at nearby
    frequencies; 
    consequently, the data in these narrow bands
    provide a high signal-to-noise estimate
    of each detector's response to atmospheric
    signal.
    The narrowband atmospheric noise features
    are completely removed using this modified
    S10 algorithm, and the amount of residual
    broadband atmospheric noise is slightly
    reduced compared to the results
    from the original S10 algorithm.

\begin{figure}
  \plotone{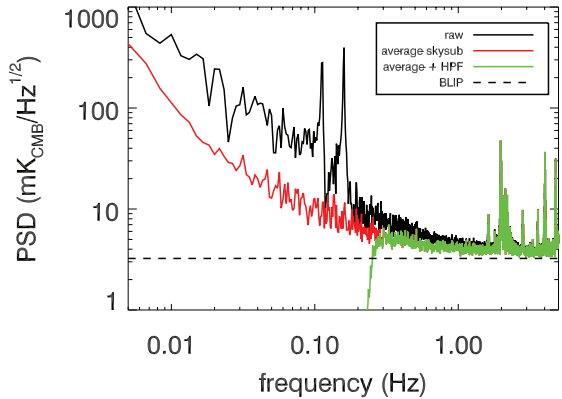}
  \caption{Timestream noise PSD for a typical
    Bolocam detector. The black curve shows the
    raw PSD recorded by the detector;
    spectral lines at the fundamental scan frequencies
    are clearly seen above the broadband
    atmospheric noise.
    The red curve shows the noise PSD after
    subtracting the atmospheric noise
    using the average signal over the FOV.
    This timestream is then high-pass filtered
    at 250~mHz to produce the green PSD.
    Note that there is very little cluster
    signal above $\simeq 2$~Hz, where there
    are some spectral lines due to 
    the readout
    electronics.
    The dashed horizontal line provides an
    estimate of the photon, or BLIP,
    noise.}
  \label{fig:psd}
\end{figure}
    
    After applying this average subtraction
    algorithm to the timestream data, we then
    high-pass filter the data according to
    \begin{displaymath}
     \mathcal{F} = 1 - \frac{1}{1 + (10^{f/f_0 - 1})^\kappa} 
    \end{displaymath}    
    with $f_0 = 250$~mHz and $\kappa = 8$.
    The value of $f_0$ was chosen to maximize
    the spatially-extended S/N for the typical cluster in our
    sample based on tests with 
    $f_0$ varying from 
    0 to 400~mHz,
    and the value of $\kappa$ was chosen to 
    produce a sharp cutoff with minimal ringing.
    Figure~\ref{fig:psd} shows a typical pre and
    post-subtraction timestream noise PSD.

  \subsection{Transfer function of the atmospheric noise filtering}
    \label{sec:xfer}

    In addition to subtracting atmospheric noise,
    the FOV-average subtraction and timestream
    high-pass filter also remove some cluster signal.
    Since we use the data timestreams to both determine
    the atmospheric fluctuation template and the
    correlation coefficient of each detector's data
    timestream with the template, the FOV-average
    subtraction acts on the data in a non-linear way.
    Consequently, 
    its impact on the data depends on
    the cluster signal.
    As described below, we quantify the effects
    of the FOV-average subtraction and the
    timestream high-pass filter via simulation by
    processing a known cluster image through
    our data-reduction pipeline.
    These simulations are computation-time
    intensive, and 
    we find, in practice (see Section~\ref{sec:tests_model}),
    that the filtering is only mildly dependent
    on the cluster signal.
    Thus, in the end, we determine the 
    effects of the filtering for a particular
    cluster's data set using the cluster
    model that best fits those data.
    We use the term transfer function to describe
    the effect of the filtering,
    although this terminology is not rigorously
    correct because the filtering depends
    in the cluster signal.

\begin{figure*}
  \plottwo{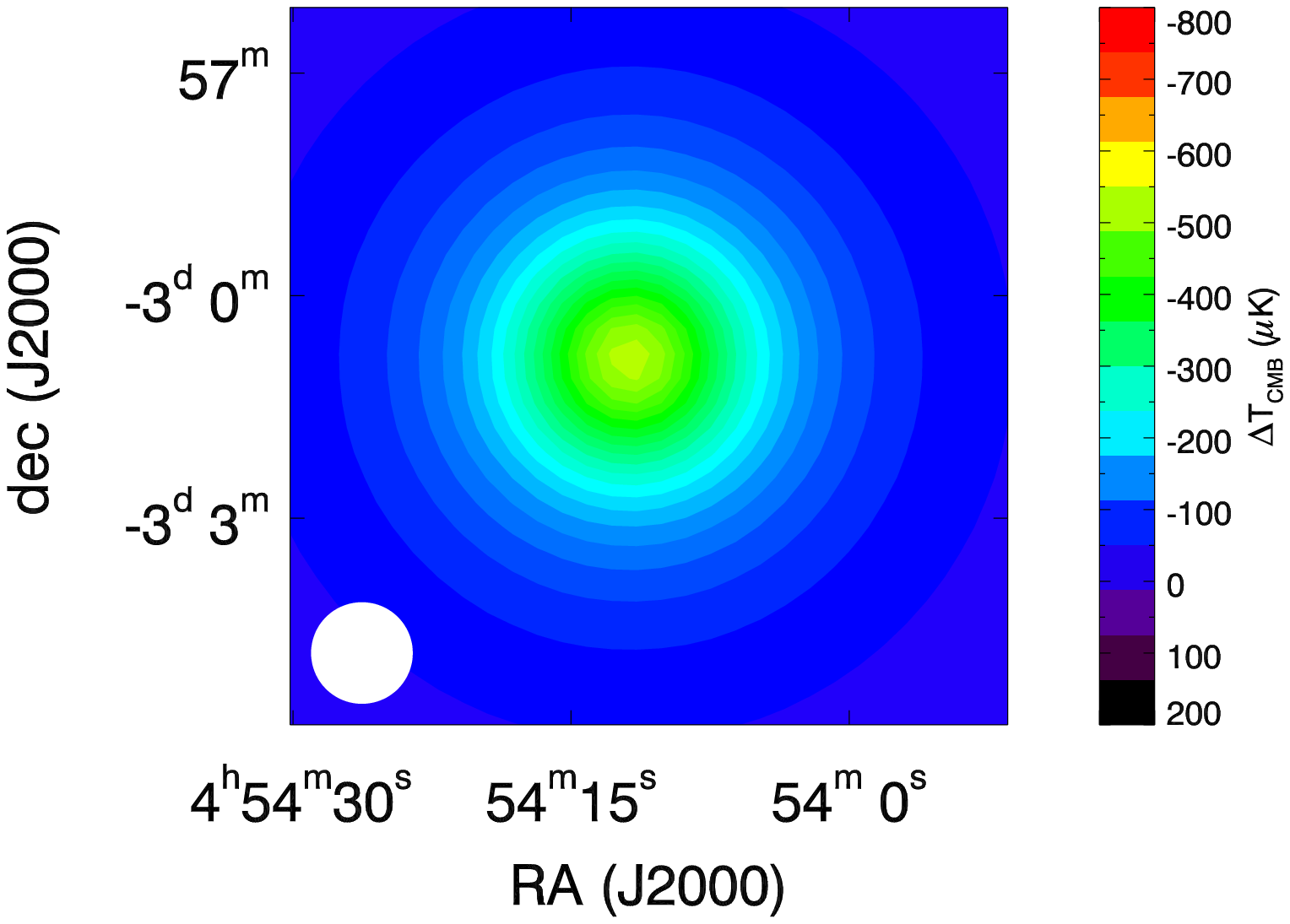}{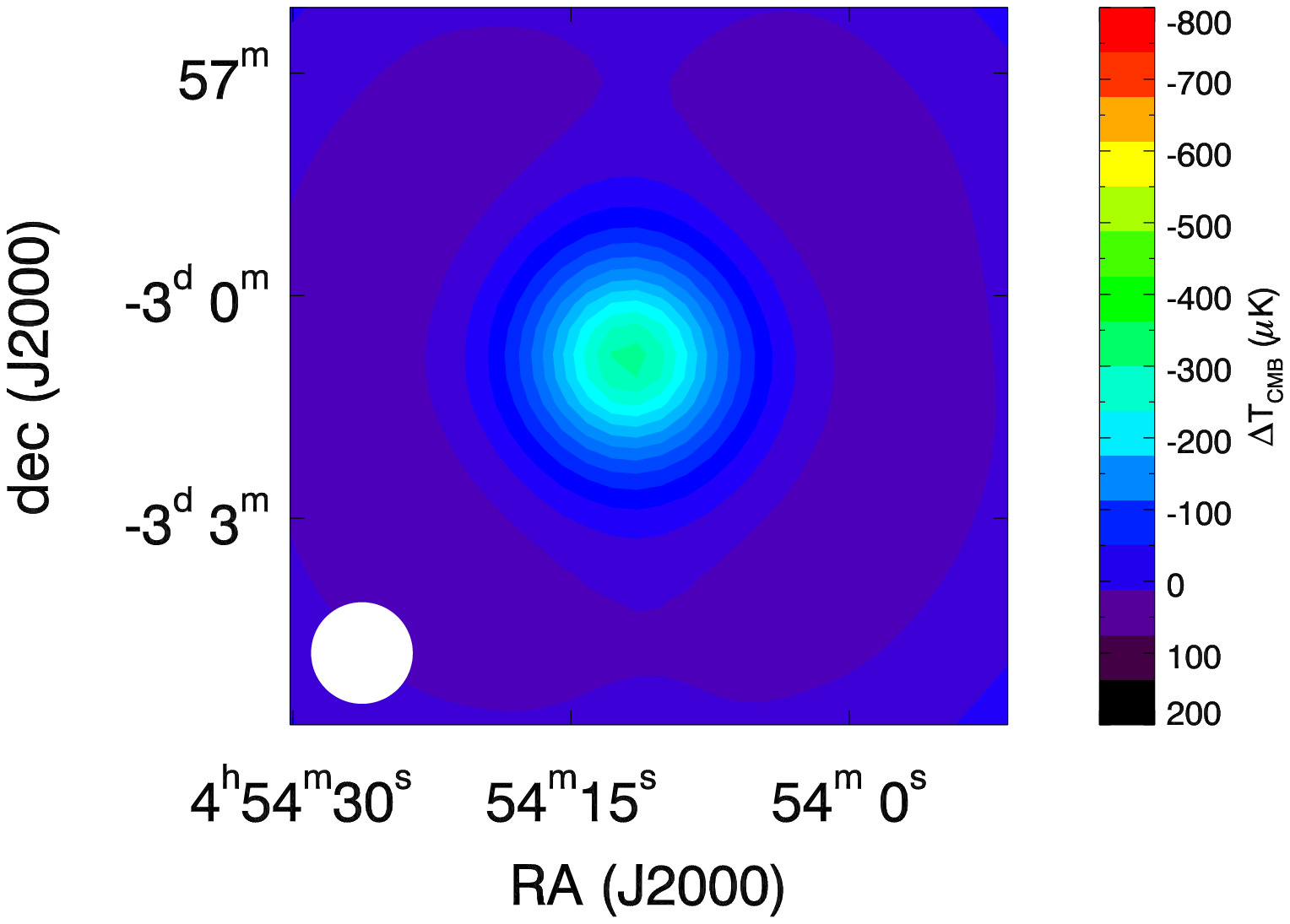}
  \caption{Images of the best-fit spherical Nagai model
    for MS 0451.6-0305.
    The left image is the model and the right image
    is the model after being processed through our
    data reduction pipeline,
    which high-pass filters the image in a 
    complex way.
    This filtering significantly reduces
    the peak decrement of the cluster and
    creates a ring of positive flux
    at $r \gtrsim 2$~arcmin.
    Note that the processed image is not
    quite azimuthally symmetric.}
  \label{fig:xfer}
\end{figure*}

\begin{figure}
  \plotone{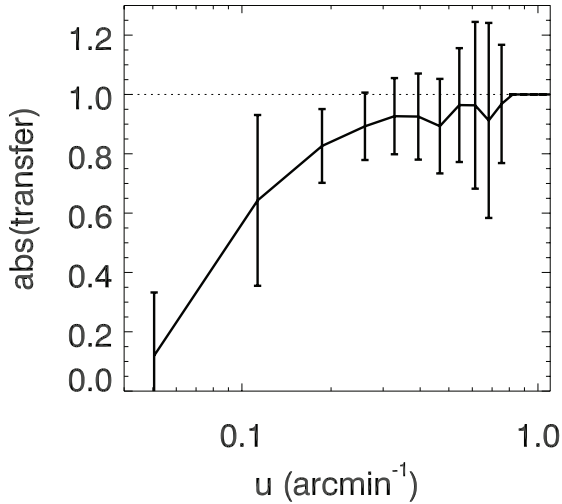}
  \caption{The magnitude of the transfer
    function for MS 0451.6-0305 as a 
    function of Fourier wavenumber $u = 1/\lambda$.
    At large scales, or small $u$,
    the measurement error is negligible and
    the error bars provide an indication
    of the azimuthal variation.
    At $u > 0.75$~arcmin$^{-1}$, the measurement
    error becomes non-negligible and 
    we set the transfer function equal 
    to 1.
    Note that this azimuthally averaged
    transfer function is for display purposes only;
    we have used the full two-dimensional transfer
    function throughout our analysis.}
  \label{fig:xfer2}
\end{figure}

    To compute the transfer function, we first insert
    a simulated, beam-smoothed cluster profile into our 
    data timestreams by reverse mapping it
    using our pointing information.
    These data are then processed in an identical
    way to the original data, and an output image,
    or map, is produced.
    When processing the data-plus-simulated-cluster
    timestreams we use the FOV-average subtraction
    correlation coefficients that were determined
    for the original data.
    This ensures that the simulated cluster is 
    processed in an identical way to the real
    cluster in our data.
      In the limit that the best-fit cluster model
      is an accurate description of the data,
      this process is rigorously correct.
    The original data map is then subtracted from
    this data-plus-simulated-cluster map to
    produce a noise-free image of the processed
    cluster.
    In Figure~\ref{fig:xfer}, we show an example
    cluster image, along with the 
    noise-free processed image of the same cluster.
    The Fourier transform of this
    processed cluster image is divided by the Fourier transform
    of the input cluster to determine how
    the cluster is filtered as a function
    of 2-dimensional Fourier mode (\emph{i.e.}, 
    what we term the 
    transfer function, see Figure~\ref{fig:xfer2}).  

    At small angular scales, there is very little
    signal in the beam-smoothed input cluster,
    and numerical noise prevents us from
    accurately characterizing the transfer
    function at these scales.
    The transfer function is expected to be
    unity at small angular scales, and is asymptoting
    to this value at the larger scales where
    we can accurately characterize it.
    Therefore, we set the transfer function to
    a value of 1 for $u > 0.75$~arcmin$^{-1}$.
    Deconvolving the processed cluster image
    using the transfer function
    (see Section~\ref{sec:images}), which
    has been approximated as 1 at small angular
    scales, produces an image that is slightly
    biased compared to the input cluster.
    The residuals between these two images are
    approximately white, with an RMS of $\lesssim 0.1$~\uK.
    This transfer-function-induced bias
    is negligible compared to our 
    noise, which has an RMS of $\simeq 10$~\uK.

    As noted above, the transfer function (weakly) depends 
    on the
    profile of the cluster;
    larger clusters are more heavily filtered than
    smaller clusters.
    Therefore, we determine a unique transfer 
    function for each cluster using the 
    best-fit elliptical Nagai model
    for that cluster~\citep[hereafter N07]{nagai07}.
    The details of this fit are given in 
    Section~\ref{sec:cluster_model}.
    Since the transfer function depends
    on the best-fit model, and vice versa,
    we determine the best-fit model
    and transfer function in an iterative way.
    Starting with a generic cluster profile, we 
    first determine a transfer function,
    and then fit an
    elliptical Nagai model using this transfer function
    (\emph{i.e.}, the Nagai model parameters
    are varied while the transfer function is 
    held fixed).
    This process is repeated, using the
    best-fit model from the previous iteration
    to calculate the transfer function,
    until the best-fit model parameters
    stabilize.
    This process converges fairly quickly,
    usually after a single iteration
    for the clusters in our sample.
    The model dependence of the resulting
    transfer function is quantified in
    Section~\ref{sec:tests_model}.

  \subsection{Noise estimation}
    \label{sec:noise_est}
 
\begin{figure}
  \plotone{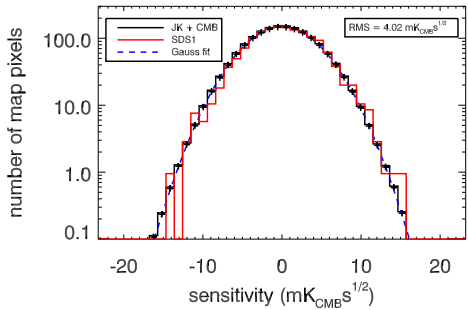}
  \caption{Histogram of the per-pixel sensitivity
    for our maps of the blank-field SDS1.
    The black line shows the average histogram
    for each of the 1000 jackknife realizations
    of the data, with noise from the CMB and 
    unresolved point 
    sources added as described in Section~\ref{sec:noise_est}.
    The sensitivities derived from the
    noise realizations are well described by a 
    Gaussian fit, with the fit quality
    quantified by a PTE of 0.44.
    Overlaid in red is the histogram for our map
    of SDS1, which is also well described
    by the Gaussian fit to the noise
    realizations (PTE = 0.63),
    indicating that our noise model
    adequately describes the data.
    The dashed blue line shows the best-fit
    Gaussian to the noise realizations and
    has an RMS of 4.02~\uK-s$^{1/2}$.}
  \label{fig:sens_hist}
\end{figure}

    In order to accurately characterize the 
    sensitivity of our images,
    we compute our map-space noise
    directly from the data via 1000
    jackknife realizations of our cluster images.
    In each realization, random subsets of half
    of the $\simeq 100$ observations
    are multiplied by $-1$ prior
    to adding them into the map.
    Each jackknife preserves the noise properties
    of the map while removing all of the astronomical
    signal, along with any possible fixed-pattern
    or scan-synchronous noise due to the 
    telescope scanning motion\footnote{
      We show that there is no measurable
      fixed-pattern or scan-synchronous noise
      in our data later in this section
      and in Section~\ref{sec:tests_noise}.}.
    Since these jackknife realizations remove
    all astronomical signal, we estimate
    the amount of astronomical noise in our images
    separately, as described below.
    After normalizing the noise estimate of each
    map pixel in each jackknife by the square root
    of the integration time in that pixel,
    we construct a sensitivity histogram, in \uK-s$^{1/2}$,
    from the ensemble of map pixels in 
    all 1000 jackknifes.
    The width of this histogram provides an
    accurate estimate of our map-space sensitivity
    (see Figure~\ref{fig:sens_hist}).
    We then assume that the noise covariance matrix
    is diagonal\footnote{
      This approximation is justified for
      our processed data maps in Section~\ref{sec:tests_noise}.
      Note that the approximation fails
      for our deconvolved images
      (see Section~\ref{sec:images}), which contain
      a non-negligible amount of correlated
      noise.
      We describe how this correlated noise
      is accounted for in our results in Section~\ref{sec:images}.
      }
    and divide by the square root of the integration time in each
    map pixel to determine the noise RMS in 
    that pixel.
    This method is analogous to the one used in 
    S09, where it is 
    described in more detail.

    There is a non-negligible amount of noise
    in our maps from two main types of astronomical
    sources:
    anisotropies in the CMB and unresolved point
    sources.
    The South Pole Telescope (SPT) has recently published
    power spectra for both of these sources
    at 150~GHz
    over the range of angular scales probed
    by our maps, so we use their measurements
    to estimate the astronomical noise in our
    maps~\citep{lueker10, hall10}.
    Note that these measurements inherently include all
    astrophysical effects, such as lensing and
    clustering of point sources.
    Using the SPT power spectra,
    we generate simulated maps of our cluster
    fields assuming Gaussian fluctuations.
    Note that this is a poor assumption for both the SZ-sourced
    CMB fluctuations and the signal sourced
    by background galaxies that are lensed by the cluster.
    However, since the noise power from both of these
    sources is quasi-negligible compared
    to the total noise in our images,
    any failure of our Gaussian assumption
    will have a minimal impact on our results.
    The good match of the 
    SDS1 data to our noise model validates
    our Gaussian assumption (see Figure~\ref{fig:sens_hist}).
    These simulated images of the CMB plus point sources
    are then reverse mapped
    into our timestream data and processed to
    estimate how they will appear in our
    cluster maps.
    We then add these processed astronomical
    realizations to our jackknife realizations
    to provide a complete estimate of the noise
    in our images.
    Sensitivity estimates from our signal-free
    map of SDS1 agree well with the sensitivity
    estimates from our noise maps,
    indicating that there are no additional
    noise sources that have not been included
    in our estimate
    (see Figure~\ref{fig:sens_hist}).
    The properties of this noise are described 
    in more detail in Section~\ref{sec:tests_noise}.
    In general, the noise from astronomical
    sources is quasi-negligible compared
    to the other noise in our images.
    However, the large-scale correlations
    from CMB fluctuations are non-negligible
    in our deconvolved images (see Table~\ref{tab:y_sz}).

    Note that we have chosen to
    model the signal
    from point sources as an additional noise
    term in our maps
    rather than attempting to subtract
    any individual point sources.
    Part of the motivation for this approach
    is the fact that we do not detect any point
    sources in our cluster or blank-field images.
    For reference, P10 mapped 15 clusters
    to a similar depth using the SPT and only 
    detected two point sources;
    the combined area of the SPT images is approximately
    20 times the combined area of our 6 images.
    Although there are several known SMGs and AGNs
    in our images~\citep{zemcov07, ivison00, cooray98},
    even the brightest sources
    will have a flux of $\simeq 10$~\uK~in
    our observing band,
    rendering them undetectable
    given our noise RMS of $\simeq 10$~\uK.
    Additionally, it is not possible to
    reliably estimate
    the flux of these sources in our
    observing band since most have only been
    detected at one or two wavelengths,
    generally separated from our
    observing band by more than a factor of 2 in 
    wavelength.

\section{Cluster model fitting}
  \label{sec:cluster_model}

  \subsection{The SZ effect}

  As mentioned above, the SZ effect involves
  CMB photons inverse Compton scattering
  off of hot electrons in the ICM.
  Since the electrons are many orders of 
  magnitude hotter than the CMB photons,
  there is, on average, a net increase in the energy
  of the photons.
  The classical distortion relative to the blackbody
  spectrum of the CMB is given by
  \begin{displaymath}
   f(x) =  x \frac{e^x + 1}{e^x - 1} - 4,
  \end{displaymath}
  where $x = h \nu / k_B T_{CMB}$.
  The magnitude of the distortion is 
  proportional to the product of the 
  density and temperature of the electrons in the ICM
  projected along the line of sight.
  The frequency-dependent temperature change
  of
  the CMB is given by
  \begin{displaymath}
    T_{SZ} = f(x) y T_{CMB},
  \end{displaymath}
  where 
  \begin{displaymath}
    y = \int n_e \sigma_T \frac{k_B T_e}{m_e c^2},
  \end{displaymath}
  and $n_e$ and  $T_e$ are the density
  and temperature of the ICM electrons.
  Relativistic corrections can be included
  by multiplying $f(x)$ by 
  $(1 + \delta(x, T_e))$~\citep{itoh98}.
  Note that the X-ray brightness of the ICM
  is proportional to $n_e^2 T_e^{1/2}$;
  the differing density and temperature
  dependence of the SZ and X-ray signals 
  makes them
  highly complementary probes of the 
  ICM~(\emph{e.g.}, \citet{bonamente06}, hereafter B06).

  \subsection{Models}
  
    We have fit our data to
    two types of models: an isothermal beta 
    model~\citep{cavaliere76, cavaliere78}
    and the pressure profile proposed by N07,
    hereafter the Nagai profile.
    The isothermal beta model has been used extensively
    to describe X-ray and SZ measurements of the 
    ICM, and our beta model fit parameters
    can be directly compared to these
    previous results.
    However, the beta model provides a poor 
    description of deep X-ray data,
    which generally have a cuspier core and 
    steeper outer profile
    (\emph{e.g.}, \citet{vikhlinin06}, N07,  and \citet{arnaud09}).
    In contrast, the Nagai pressure profile,
    which is a generalization
    of the NFW dark matter profile, is able to
    describe deep X-ray observations over a wide
    range of scales (N07, \citet{arnaud09}).
    Therefore, we have also fit our data
    to the Nagai model.
    In practice, the beta and Nagai
    models are highly degenerate over the
    range of angular scales to which our
    data are sensitive ($1-12$~arcmin).

    Specifically, the isothermal beta model is described by
    \begin{displaymath}
      p = \frac{p_0}{(1 + r^2/r_c^2)^{3 \beta/2}},
    \end{displaymath}
    where $p$ is the pressure profile, $p_0$ is the 
    pressure normalization, $r_c$ is the core
    radius, and $\beta$ is the power law slope.
    The beta model can be analytically integrated along
    the line of sight to give the observed SZ signal,
    with
    \begin{displaymath}
       T_{SZ} = 
      \frac{f(x) y_0 T_{CMB}}
       {(1 + r^2/r_c^2)^{(3\beta-1)/2}} + \delta T,
   \end{displaymath}
    where $T_{SZ}$ is the SZ signal in our map (in \uK)
    and $y_0$ is the central Comptonization.
    We fix the value of $\beta$ at 0.86 for all of our
    fits;
    this is the best-fit value found in P10
    for SZ data\footnote{
      As P10 point out,
      this value for $\beta$ is larger than those
      generally found from X-ray data, 
      in agreement with simulations~\citep{hallman07}.
      Additionally, it suggests that the ICM 
      temperature is falling with increasing
      radius, in agreement with simulations and 
      data (\emph{e.g.}, N07).}.
    Since our data timestreams are high-pass filtered, we
    are not sensitive to the DC signal level
    in our images.
    Note that because the timestream data, rather than the map,
    are high-pass filtered,
    the DC signal level of the map is in general
    not equal to 0.
    However, the DC signal of the map 
    is not physically
    meaningful due to the filtering and must therefore
    be included as a free parameter, $\delta T$,
    in the model fits.
    We have also generalized the beta model to
    be elliptical in the plane of the sky with
    \begin{displaymath}
      T_{SZ} = 
      \frac{f(x) y_0 T_{CMB}}
       {(1 + \frac{r_1^2}{r_c^2} + \frac{r_2^2}{(1-\epsilon)^2r_c^2})^{(3\beta-1)/2}} + \delta T,
    \end{displaymath}
    where $r_1$ is oriented along the major axis, 
    described by a position angle of $\theta$
    (in degrees east of north),
    $r_2$ is orthogonal to the major axis,
    and $\epsilon$ is the ellipticity.
 
    The Nagai model is described by
    \begin{equation}
      p = \frac{p_0}{(r/r_s)^{\mathcal{C}} \left[1 + 
      (r/r_s)^\mathcal{C}\right]^{(\mathcal{B}
      -\mathcal{C})/\mathcal{A}}},
    \end{equation}
    where $p$ is the pressure,
    $p_0$ is the pressure normalization, $r_s$ is the
    scale radius ($r_s = r_{500}/c$, with
    $c \simeq 1.2$~\citep{arnaud09}), 
    and $\mathcal{A}$, $\mathcal{B}$, and $\mathcal{C}$
    are the power law slopes at intermediate, large, 
    and small radii compared to $r_s$.
    We have fixed the values of $\mathcal{A}$, $\mathcal{B}$, 
    and $\mathcal{C}$ to the best-fit values found 
    in \citet{arnaud09} (1.05,5.49,0.31).
    The Nagai model is not analytically integrable,
    so we numerically integrate $p$ along the line of
    sight to determine $T_{SZ}$.
    We have also generalized the Nagai model to be
    elliptical in the plane of the sky, 
    using the same notation as our elliptical
    generalization of the beta model.
    Note that, in all of our fits, we have corrected 
    for relativistic effects via the approximations
    given in \citet{itoh98} using gas temperature
    estimates from X-ray observations of these 
    clusters~\citep{cavagnolo08, jeltema01}.
    The relativistic
    corrections are $\simeq 5$\% for the typical
    temperatures in these massive clusters ($\simeq 10$~keV).

    We find that both the beta model and the Nagai
    model adequately describe our data,
    with the exception of Abell 697,
    and neither model is preferred with
    any significance.
    However, the elliptical models ($\chi^2$/DOF $ = 5605/5413$) 
    provide a much better fit to the full ensemble
    of our cluster data compared to the
    spherical models ($\chi^2$/DOF $ = 5813/5423$, 
    see Tables~\ref{tab:nagai_model} and \ref{tab:beta_model}).
    The $F$-ratio for these two fits is 20.1 for
    $\nu_1 = 10$ and $\nu_2 = 5413$ degrees of freedom
    \citep{bevington92},
    corresponding to a probability of $< 10^{-36}$
    that we would obtain data with equal or greater preference
    for elliptical models if the clusters were spherical.
    If we neglect Abell 697, which is not well described by
    either model, then the $F$-ratio is 12.2 with
    a corresponding probability of $< 10^{-16}$.
    The weighted mean ellipticity of the five
    clusters is $\epsilon = 0.27 \pm 0.03$, 
    consistent with results from X-ray data
    (\emph{e.g.}, \citet{maughan08, defilippis05}).
    Our inability to distinguish between the Nagai
    and beta models is likely due to the limited 
    spatial dynamic range of our images
    ($\simeq 1 - 12$~arcmin).
    For the clusters in our sample, we are insensitive
    to the core, $r \lesssim 0.15r_{500} \simeq 0.5$~arcmin, and the 
    outskirts of the cluster $r \gtrsim 2r_{500} \simeq 10$~arcmin,
    which means we are only sensitive to a single power law
    exponent in the Nagai model, $\mathcal{A}$.
    For this reason, the Nagai model is highly degenerate
    with the beta model for the angular scales probed 
    by our data.

\begin{deluxetable*}{ccccccccc} 
  \tablewidth{0pt}
  \tablecaption{Nagai model fit parameters}
  \tablehead{\colhead{cluster} & \colhead{$p_0$ ($10^{-11} \frac{\textrm{erg}}{\textrm{cm}^3}$)} &
    \colhead{$r_s$ (arcmin)} & \colhead{$c_{500}$} &
    \colhead{$\epsilon$} & 
    \colhead{$\theta$ (deg)} & 
    \colhead{$\chi^2$/DOF} & \colhead{PTE$_{\chi^2}$} &
    \colhead{PTE$_{\textrm{sim}}$}}
  \startdata
    \sidehead{elliptical Nagai model}
    Abell 697 & \phn$9.3 \pm 1.3$ & $6.9 \pm 1.0$ & $0.93 \pm 0.14$ &
      $0.37 \pm 0.05$ & $-24 \pm 4$ & 
      1289/1117 & 0.00 & 0.00 \\
    Abell 1835 & \phn$8.1 \pm 1.7$ & $6.7 \pm 1.5$ & $0.94 \pm 0.21$ &
      $0.27 \pm 0.07$ & \phn$-16 \pm 10$ & 
      966/945 & 0.31 & 0.22 \\
    MS 0015.9+1609 & \phn$6.7 \pm 1.4$ & $5.5 \pm 1.1$ & $0.62 \pm 0.13$ &
      $0.24 \pm 0.08$ & \phs\phn$68 \pm 12$ & 
      1079/1117 & 0.79 & 0.73 \\
    MS 0451.6-0305 & \phn$8.7 \pm 1.6$ & $4.7 \pm 0.7$ & $0.80 \pm 0.14$ &
      $0.26 \pm 0.06$ & \phs$85 \pm 7$ & 
      1188/1117 & 0.07 & 0.08 \\
    MS 1054.4-0321 & \phn$6.5 \pm 1.9$ & $3.6 \pm 1.0$ & $0.64 \pm 0.19$ &
      $0.09 \pm 0.07$ & \phn\phn$-1 \pm 32$ & 
      1084/1117 & 0.75 & 0.72 \\
    \sidehead{spherical Nagai model}
    Abell 697 & $11.0 \pm 1.8$ & $4.6 \pm 0.8$ & $1.40 \pm 0.22$ & 
      - & - & 
      1399/1119 & 0.00 & 0.00 \\ 
    Abell 1835 & \phn$8.9 \pm 2.1$ & $5.1 \pm 1.2$ & $1.24 \pm 0.27$ &
      - & - & 
      1007/947 & 0.08 & 0.07 \\ 
    MS 0015.9+1609 & \phn$6.0 \pm 1.1$ & $5.4 \pm 1.1$ & $0.63 \pm 0.12$ &
      - & - & 
      1100/1119 & 0.66 & 0.61 \\
    MS 0451.6-0305 & $10.6 \pm 2.1$ & $3.5 \pm 0.5$ & $1.08 \pm 0.15$ &
      - & - & 
      1220/1119 & 0.02 & 0.02 \\ 
    MS 1054.4-0321& \phn$6.0 \pm 1.8$ & $3.6 \pm 1.0$ & $0.64 \pm 0.16$ &
      - & - & 
      1087/1119 & 0.75 & 0.75 
   \enddata
  \tablecomments{Table of the best-fit parameters and 1$\sigma$
    uncertainties for our Nagai model fits
    with the power-law exponents fixed to the best-fit values found
    in \citet{arnaud09} (1.05,5.49,0.31).
    From left to right the columns give the normalization
    of the Nagai profile, $p_0$, the scale radius of the major axis, $r_s$,
    the concentration parameter, $c_{500}$, 
    the ellipticity, $\epsilon$, the position angle of the 
    major axis in degrees east of north, $\theta$,
    the $\chi^2$ and DOF for the fit, and the 
    goodness of fit quantified by the probability to
    exceed the given $\chi^2$/DOF based on the 
    standard $\chi^2$ probability distribution function
    and also
    empirically by the fraction of our 1000
    noise realizations that produce a larger
    $\chi^2$ value when a model cluster is
    added to them
    (see Section~\ref{sec:tests_noise}).
    The values of $c_{500}$ for the spherical
    fits can be compared to
    the nominal value of 1.17 
    that is given in \citet{arnaud09}
    based on spherical fits of X-ray data.
    See Section~\ref{sec:tests_noise} for a description
    of how the parameter uncertainties are calculated.
    Note that we have included the 4.3\% uncertainty
    in our flux calibration in the error estimates for
    $p_0$.}
  \label{tab:nagai_model}
\end{deluxetable*}

\begin{deluxetable*}{ccccccccc} 
  \tablewidth{0pt}
  \tablecaption{Beta model fit parameters}
  \tablehead{\colhead{cluster} & \colhead{$y_0$ ($10^{-4}$)} &
    \colhead{$r_c$ (arcsec)} & \colhead{$r_c/r_{500}$} &
    \colhead{$\epsilon$} & 
    \colhead{$\theta$ (deg)} & 
    \colhead{$\chi^2$/DOF} & \colhead{PTE$_{\chi^2}$} &
    \colhead{PTE$_{\textrm{sim}}$}}
  \startdata
    \sidehead{elliptical beta model}
    Abell 697 & $2.93 \pm 0.26$ & \phn$98 \pm 10$ & $0.26 \pm 0.03$ &
      $0.38 \pm 0.04$ & $-10 \pm 2$ & 
      1288/1117 & 0.00 & 0.00 \\ 
    Abell 1835 & $2.49 \pm 0.31$ & \phn$79 \pm 11$ & $0.21 \pm 0.03$ &
      $0.28 \pm 0.07$ & \phn\phn$-9 \pm 10$ & 
      970/945 & 0.28 & 0.20 \\
    MS 0015.9+1609 & $2.59 \pm 0.27$ & \phn$85 \pm 12$ & $0.43 \pm 0.07$ &
      $0.24 \pm 0.07$ & \phs\phn$70 \pm 12$ & 
      1078/1117 & 0.79 & 0.73 \\
    MS 0451.6-0305 & $2.89 \pm 0.22$ & $68 \pm 8$ & $0.31 \pm 0.04$ &
      $0.26 \pm 0.06$ & \phs$80 \pm 8$ & 
      1193/1117 & 0.06 & 0.06 \\  
    MS 1054.4-0321 & $2.13 \pm 0.22$ & \phn$58 \pm 12$ & $0.41 \pm 0.10$ &
      $0.09 \pm 0.07$ & \phs\phn\phn$7 \pm 33$ & 
      1083/1117 & 0.76 & 0.73 \\  
    \sidehead{spherical beta model}
    Abell 697 & $2.74 \pm 0.24$ & $72 \pm 8$ & $0.19 \pm 0.02$ &
      - & - & 
      1397/1119 & 0.00 & 0.00 \\ 
    Abell 1835 & $2.46 \pm 0.31$ & $63 \pm 8$ & $0.17 \pm 0.02$ &
      - & - & 
      1008/947 & 0.08 & 0.08 \\
    MS 0015.9+1609 & $2.54 \pm 0.29$ & \phn$80 \pm 11$ & $0.40 \pm 0.05$ &
      - & - & 
      1099/1119 & 0.66 & 0.61 \\
    MS 0451.6-0305 & $2.92 \pm 0.24$ & $53 \pm 7$ & $0.24 \pm 0.03$ &
      - & - & 
      1222/1119 & 0.02 & 0.02 \\  
    MS 1054.4-0321 & $2.13 \pm 0.22$ & \phn$56 \pm 10$ & $0.40 \pm 0.07$ &
      - & - & 
      1085/1119 & 0.76 & 0.76 
   \enddata
  \tablecomments{Table of the best-fit parameters and 1$\sigma$
    uncertainties for our beta model fits
    with the power law exponent $\beta$ set
    to the best-fit value found in P10 (0.86).
    From left to right the columns give the normalization
    of the beta profile, $y_0$, the core radius of the major axis, $r_c$,
    the relative value of the core radius, $r_c/r_{500}$, 
    the ellipticity, $\epsilon$, the position angle of the 
    major axis in degrees east of north, $\theta$,
    the $\chi^2$ and DOF for the fit, and the 
    goodness of fit quantified by the probability to
    exceed the given $\chi^2$/DOF based on the 
    standard $\chi^2$ probability distribution function
    and also
    empirically by the fraction of our 1000
    noise realizations that produce a larger
    $\chi^2$ value when a model cluster is
    added to them
    (see Section~\ref{sec:tests_noise}).
    The values of $r_c/r_{500}$ for the spherical fits
    can be compared to
    the nominal value of 0.20 given in P10
    based on spherical fits of SPT SZ data.
    See Section~\ref{sec:tests_noise} for a description
    of how the parameter uncertainties are calculated.
    Note that we have included the 4.3\% uncertainty
    in our flux calibration in the error estimates for
    $y_0$.}
  \label{tab:beta_model}
\end{deluxetable*}

\begin{deluxetable*}{cccccccc} 
  \tablewidth{0pt}
  \tablecaption{Beta model fit parameters compared to OVRO/BIMA/\emph{Chandra} results}
  \tablehead{\colhead{cluster} & \multicolumn{2}{c}{$y_{0}$ ($10^{-4}$)} &
    \colhead{$\beta$} & \multicolumn{2}{c}{$r_c$ (arcsec)} &
    \colhead{$\delta$RA} & 
    \colhead{$\delta$dec} \\
    \colhead{} & \colhead{Bolocam} & \colhead{OV/BI/\emph{Ch}} &
    \colhead{} & \colhead{Bolocam} & \colhead{OV/BI/\emph{Ch}} & \colhead{} & \colhead{}
} 
  \startdata
    Abell 697 & $3.63 \pm 0.31$ & $2.29^{+0.23}_{-0.24}$ & 0.607 & $49 \pm 6$ & 
      $43.2^{+2.1}_{-2.0}$ & \phn$-6.2 \pm 4.2$ & $-14.0 \pm 4.6$ \\
    Abell 1835 & $2.95 \pm 0.37$ & $3.19^{+0.19}_{-0.21}$ & 0.670 & $49 \pm 6$ & 
      $32.4^{+1.4}_{-1.1}$ & \phn$-8.2 \pm 4.4$ & \phs\phn$1.5 \pm 5.0$ \\
    MS 0015.9+1609 & $2.80 \pm 0.31$ & $2.55^{+0.15}_{-0.15}$ & 0.744 & \phn$69 \pm 10$ & 
      $42.9^{+2.6}_{-2.4}$ & \phs$18.2 \pm 5.2$ & \phn\phs$5.3 \pm 4.7$ \\
    MS 0451.6-0305 & $3.05 \pm 0.25$ & $2.72^{+0.15}_{-0.13}$ & 0.795 & $48 \pm 6$ & 
      $36.0^{+1.9}_{-1.6}$ & \phs$14.2 \pm 4.6$ & \phn\phs$7.9 \pm 4.2$ \\
    MS 1054.4-0321 & $1.92 \pm 0.20$ & $2.09^{+0.17}_{-0.17}$ & 1.083 & \phn$70 \pm 12$ & 
      $70.5^{+6.5}_{-6.9}$ & \phn$-1.0 \pm 4.4$ & \phn$-1.5 \pm 4.6$ 
   \enddata
  \tablecomments{Table of the best-fit parameters and 1$\sigma$ uncertainties
    when we fit a spherical beta model to our data using the value of 
    $\beta$ found by B06 using OVRO/BIMA and \emph{Chandra} data
    (in contrast to our nominal beta model
    fits with $\beta = 0.86$).
    From left to right the columns give our best-fit values of $y_0$,
    the B06 best-fit values of $y_0$, the value of $\beta$,
    our best-fit values of $r_c$, the B06 best-fit values
    of $r_c$, and the RA and dec offsets of the 
    centroids of our best-fit models
    compared to the X-ray centroids of the B06 fits.
    Compared to B06, we find a significantly larger
    value of $y_0$ for Abell 697, and we
    find significantly larger values of $r_c$
    for Abell 1835 and MS 0015.9+1609.
    We also find small, but measurable, centroid
  offsets for all of the clusters other than
  MS 1054.4-0321.}
  \label{tab:bonamente}
\end{deluxetable*}

  \subsection{Fitting procedure}

    Our cluster images are filtered by both the
    atmospheric noise subtraction and the 
    Bolocam point-spread function.
    Therefore, prior to fitting a model to our
    data, we need to filter the cluster model in an
    identical way.
    First, we generate a 2-dimensional image 
    using the cluster model,
    computed directly from the isothermal
    beta model and via a line-of-sight integration
    of the Nagai pressure model.
    Next, the model image is convolved with the
    Bolocam point-spread function and the
    measured transfer function.
    In practice, the transfer function convolution
    is performed via multiplication in Fourier space.
    This filtered model is then compared to our
    data map, using all of the map pixels contained
    within a radius where the minimum coverage
    is greater than 25\% of the peak coverage.
    This radius is typically between 6 and 7~arcmin.
    We use an iterative least-squares technique to
    determine the best-fit parameters for each 
    model;
    we estimate the uncertainty on each parameter
    via the standard deviation
    of the best-fit parameter
    values estimated from noise realizations
    with model clusters added to them
    (see Section~\ref{sec:tests_noise}).

    For each model, we fit both a scale radius
    ($r_c$ for the beta model and $r_s$
    for the Nagai model) and a normalization
    ($y_0$ for the beta model and $p_0$
    for the Nagai model).
    Additionally, as described above,
    we fit for the observationally unconstrained
    DC signal level of the map, $\delta T$.
    We also fit for a centroid offset 
    relative to the X-ray pointing center.
    The offsets for the 
    five clusters range from $\simeq 0-20 \pm 5$~arcsec,
    indicating there are no major differences
    in the X-ray and SZ centroids.
    Finally, when we allow the model to be
    elliptical in the image plane, we fit
    for the ellipticity $\epsilon$ and 
    position angle $\theta$.
    A complete list of the best-fit parameters for 
    spherical and elliptical versions of both models
    is given in Tables~\ref{tab:nagai_model}
    and \ref{tab:beta_model}.

  \subsection{Discussion and comparison to previous results}
    \label{sec:comparison}

    All of these clusters have been studied extensively
    at a wide range of wavelengths,
    providing us with a large number of published
    results to compare our model fit parameters to.
    In general, our results agree well with
    those found from previous studies.
    In particular, when we fit a spherical
    beta model to our data using the values of 
    $\beta$ given in
    B06, which were 
    determined using \emph{Chandra}
    X-ray data and 30~GHz interferometric
    SZ data,
    our best-fit values
    for $y_0$ agree quite well, with
    the exception of Abell 697
    (see Table~\ref{tab:bonamente}).
    The best-fit values we find for $r_c$
    are also consistent with those determined
    in B06, with the 
    exception of Abell 1835 and MS 0451.6-0305.
    Additionally, the ellipticity and orientation
    of our elliptical fits are in general
    consistent
    with previously published results
    ~\citep{maughan08, donahue03, defilippis05,
    piffaretti03, girardi06, mcnamara06,
    schmidt01, neumann00}.
    Finally, the best-fit concentration
    parameters from our Nagai model fits 
    to three of the five clusters are
    consistent with those found from
    X-ray data~\citep{arnaud09};
    MS 0015.9+1609 and MS 1054.4-0321
    have significantly lower values of $c_{500}$.
    We describe each cluster in detail below:

  \begin{description}
    \item[Abell 697]{We find that Abell 697 has a significant
      ellipticity, and it is not well described by either
      a beta model or a Nagai model.
      The poor model fits result from
      the cluster appearing significantly extended in 
      the SW direction, and extremely compact in the 
      NE direction.
      Abell 697 is the only cluster in our sample
      that is not adequately described by the 
      models.
      The ellipticity we find for Abell 697,
      $\epsilon = 0.37 \pm 0.05$, is roughly
      consistent
      with the ellipticity of $\simeq 0.25$ found from X-ray data
      \citep{defilippis05, maughan08, girardi06}.
      We find a position angle of $-24$~deg for Abell 697,
      which is similar to the value of $-16$~deg found by
      \citet{girardi06}, but somewhat misaligned
      to the position angle of 16~deg found 
      by \citet{defilippis05}.
    }
    \item[Abell 1835]{We detect an ellipticity
      in our image of 
      Abell 1835,
      and we find that it is well described by
      either an elliptical beta or Nagai model.
      Our best-fit spherical
      Nagai model parameters with $\mathcal{A} = 0.9$, $\mathcal{B} = 5.0$,
      and $\mathcal{C} = 0.4$ ($p_0 = 11.1 \pm 2.6 \times 10^{-11}$~erg/cm$^3$, 
      $r_s = 4.6 \pm 1.1$~arcmin) are consistent
      with the values found in \citet{mroczkowski09}
      ($p_0 = 13.6 \times 10^{-11}$~erg/cm$^3$, $r_s = 4.3$~arcmin)
      using a combination of SZA SZ data and \emph{Chandra}
      X-ray data.
      X-ray measurements find an ellipticity of 
      $0.1-0.2$ for Abell 1835
      \citep{defilippis05, mcnamara06, schmidt01},
      consistent with the value of $\epsilon = 0.27 \pm 0.07$ we
      find with Bolocam.
      We find a position angle of $-16$~deg for Abell 697,
      which is similar to the values of 
      7, $-20$, and $-30$~deg found by
      \citet{defilippis05}, \citet{mcnamara06}, and
      \citet{schmidt01}.}
    \item[MS 0015.9+1609]{MS 0015.9+1609 appears to be
      elliptical in our image, 
      and it is well described by either a 
      spherical or elliptical model.
      As with Abell 697 and Abell 1835, X-ray data
      favor slightly lower ellipticities, 
      $\epsilon \lesssim 0.20$, compared to
      what we find with Bolocam,
      $\epsilon = 0.24 \pm 0.08$~\citep{defilippis05, maughan08, piffaretti03}.
      We find a position angle of 68~deg for 
      MS 0015.9+1609, fairly close to the value
      of 47~deg found by \citet{piffaretti03},
      but almost orthogonal to the value of 
      $-49$~deg found by \citet{defilippis05}.}
    \item[MS 0451.6-0305]{MS 0451.6-0305 also 
      appears to be elliptical in our image
      and is adequately described by either
      an elliptical beta or Nagai model.
      We find an ellipticity of $\epsilon = 0.26 \pm 0.06$
      for MS 0451.6-0305, in excellent agreement with
      ellipticities determined using X-ray data
      \citep{defilippis05, donahue03}.
      Additionally, our best-fit position angle
      of 85~deg agrees well with the value
      of 84~deg found by \citet{defilippis05} and 
      the value of $-75$~deg found by \citet{donahue03}.}
    \item[MS 1054.4-0321]{Our image of MS 1054.4-0321
      shows no evidence for ellipticity, and it
      is well described by either an elliptical
      or spherical model.
      Although MS 1054.4-321 does not appear to be 
      elliptical in our data, X-ray data
      show a clear ellipticity oriented 
      along the east-west direction
      \citep{neumann00, jeltema01}.
      However, our non-detection of an ellipticity,
      $\epsilon = 0.09 \pm 0.07$, is only
      marginally inconsistent
      with the X-ray value determined by
      \citet{neumann00}, 
      $\epsilon = 0.29$.}
    \item[SDS1]{We have attempted to fit
      cluster models to the SDS1 map, but
      we find best-fit amplitudes that are
      consistent with 0 and 
      best-fit scale radii that are 
      large compared to the size
      of our images
      (\emph{i.e.}, scale radii that
      are large enough to produce
      profiles that are approximately
      constant over the entire 
      image).}
  \end{description}

\section{Model-independent images and $Y_{SZ}$ estimates}
  \label{sec:images}

  Rather than using models, which at best provide
  an adequate description of what clusters look like
  on average, we have chosen to derive observable
  quantities from our images in a quasi-model-independent
  way.
  In Section~\ref{sec:xfer} we described how we calculate
  a unique transfer function for each cluster to quantify
  the effects of our noise filtering.
  In the Fourier space of our images,
  these transfer functions are a set of 
  two-dimensional complex numbers that
  describe how an input cluster image is filtered
  as a function of two-dimensional Fourier mode.
  We obtain an unfiltered, or deconvolved, image of the cluster
  by Fourier transforming
  our image, dividing by the two-dimensional
  complex transfer function, and then Fourier transforming
  the result back to image space\footnote{
    This method can be compared to the deconvolution method
    employed by APEX-SZ for their analysis of 
    Abell 2163 and Abell 2204~\citep{nord09,basu10}.
    They first determine what a point-like object looks
    like in their image after being filtered.
    Next, they fit this filtered point-source image
    to the map pixel with the largest S/N and
    subtract it from the image.
    The process is repeated until the map is 
    consistent with noise;
    the sum of all the unfiltered point-like images
    removed from the map
    gives the deconvolved cluster image.}.
  At the largest scales in our map, the transfer function
  has a magnitude of $\simeq 0.2$
  (see Figure~\ref{fig:xfer}), resulting in a numerically stable,
  but significant, amplification of the large scale
  noise.
  In particular, the residual atmospheric noise and
  primary CMB fluctuations, which had been filtered
  to be approximately white, produce a significant
  low-frequency noise component in our deconvolved images.
  We estimate the noise in our deconvolved images
  by deconvolving each of the 1000 noise realizations
  for each cluster.
  Since the off-diagonal elements of the noise
  covariance matrix are significant due to the
  low-frequency noise, we estimate all of our measurement
  uncertainties from the standard deviation of
  measuring the same quantity in each of 
  our 1000 deconvolved 
  noise realizations (see below and Section~\ref{sec:tests_noise}).
  Note that most of the measurement uncertainties
  in our processed (\emph{i.e.}, filtered) images
  are computed in the same way, even though their
  noise covariance matrix is approximately diagonal.

  By deconvolving the two-dimensional transfer function
  of our data processing we obtain unbiased cluster
  images, modulo smoothing with our PSF and the 
  unconstrained DC signal level.
  Although the transfer function was computed using
  the best-fit-elliptical Nagai model, and is 
  therefore somewhat model dependent, the 
  dependence is negligible other than the
  determination of the DC signal offset of the image
  (see Section~\ref{sec:tests_model}).
  We do not attempt to recover the information
  lost due to smoothing by our PSF, but we 
  use the value of $\delta T$ found from our
  elliptical Nagai model fits to restore the
  correct DC signal level to our images.
  Due to uncertainties in the model itself,
  especially at large radii where there is 
  little or no observational data, along with
  cluster-to-cluster deviations from the model,
  this does introduce a non-negligible 
  model-based bias in our images
  (the typical model uncertainty in the DC
  signal offset is $5-10$~\uK,
  see Section~\ref{sec:tests_model}).
  However, we emphasize that this bias only
  affects the DC signal level of the images;
  the shapes of the cluster profiles are
  essentially model-independent.

  We have computed a model-independent 
  value for $Y_{500}$ from these
  deconvolved images, which
  is the integrated $y$ within $r_{500}$
  (\emph{i.e.}, the cylindrical $Y_{500}$
  rather than the spherical $Y_{500}$, 
  see Table~\ref{tab:y_sz}).
  As mentioned above, since our assumption that the noise covariance
  matrix is diagonal fails for the deconvolved
  images, we estimate the uncertainty in
  our estimate of $Y_{SZ}$ via the scatter
  among the $Y_{SZ}$ values determined from
  each of our noise
  realizations.
  As an example of the amount of low-frequency noise present
  in our deconvolved images on the scale of $r_{500}$, note that
  the measurement uncertainty on $Y_{500}$ is 
  approximately 10 times larger than it would be
  if the noise was white.
  For the five clusters in our sample, we are 
  able to determine $Y_{SZ}$ with an 
  uncertainty of $\simeq 10$\%,
  limited mainly by systematics in determining
  the DC signal offset and flux calibration.

\begin{figure}
  \plotone{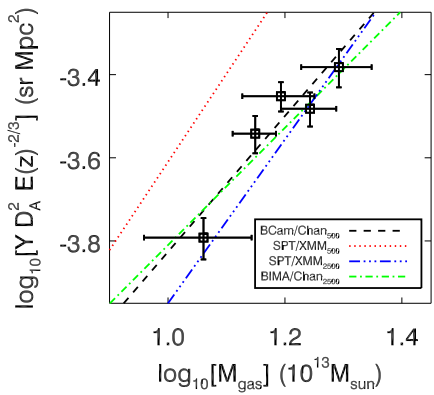}
  \caption{Scaling relation between $Y_{SZ}$ and 
    $M_{gas}$ for the five clusters in our
    sample.
    The black squares give the Bolocam
    model-independent cylindrical $Y_{500}$
    estimates
    and \emph{Chandra} $M_{gas,500}$ estimates
    (see Table~\ref{tab:clusters}),
    and the black dashed line represents the
    best fit to these data.
    We also show fits of a similar type
    obtained by various authors:
    red dashed: SPT beta-model-derived
    $Y_{500}$ vs. $M_{gas,500}$, primarily
    obtained from \emph{XMM-Newton} data (P10);
    solid blue:
    same, but within $r_{2500}$;
    dot-dashed green:
    OVRO/BIMA beta-model-derived $Y_{2500}$
    vs. $M_{gas,2500}$ obtained from
    \emph{Chandra} data \citep{bonamente08}.
    The difference between the red-dashed
    line and our data is likely caused
    by systematic differences between
    model-derived and model-independent
    estimates of $Y_{500}$\footnote{
      P10 calculated both model-derived
      and model-independent estimates of
      $Y_{SZ}$ for the 15 clusters
      in their sample.
      Within a radius of $r_{2500}$,
      the two estimates of $Y_{2500}$
      are on average consistent with
      each other.
      However, the P10 model-derived estimates
      of $Y_{500}$ are on average a factor
      of 1.5 larger than the model-independent
      estimates for the 10 clusters used
      to determine the $Y_{SZ}-M_{gas}$
      scaling relation.
      As a result, the P10 model-derived scaling
      relation for $Y_{500}-M_{gas,500}$ will
      be systematically higher by $\log_{10}(1.5) \simeq 0.2$
      compared to a model-independent scaling
      relation.}.
    The scatter of our data 
    relative to our best fit is 13\%,
    consistent with the expected intrinsic 
    scatter of $\simeq 10$\% given our
    $\simeq 10$\% uncertainty on $Y_{500}$.}
  \label{fig:ym}
\end{figure}

  We examine the self-similar scaling
  that is expected between $Y_{SZ}$ and 
  the total cluster mass using the relation
  given in \citet{bonamente08},
  \begin{displaymath}
    Y_{SZ} D_{A}^2 E(z)^{-2/3} \propto f_{gas}^{-2/3} M_{gas}^{5/3},
  \end{displaymath}
  where $D_{A}$ is the angular diameter distance,
  $E(z) = \sqrt{\Omega_m(1 + z)^3 + \Omega_{\Lambda}}$,
  and we have assumed $M_{gas}$ is a good proxy for
  the total cluster mass~\citep{allen08, mantz10}.
  The scatter in this relation is expected to be
  $\lesssim 10$\%~\citep{kravtsov06}, 
  and current measurements are roughly consistent
  with this prediction~\citep{morandi07, bonamente08, 
  marrone09, plagge10, huang10, andersson10}.
  Given our small sample, we do not attempt to
  constrain the intrinsic scatter in the
  $Y-M$ relation, but we follow the 
  formalism of \citet{marrone09} and 
  P10 to fit a logarithmic
  scaling of the form $\mathcal{Y} = 
  a + b\mathcal{X}$ with the intrinsic
  scatter set to 10\%.
  For the $Y_{500}-M_{gas,500}$ relation
  we find best-fit values of 
  $a = -5.46 \pm 0.84$ and $b = 1.63 \pm 0.71$,
  consistent with the self-similar prediction
  of $b = 5/3$.
  The overall scatter of our data about the fit
  is 13\%, consistent with an intrinsic scatter
  of $\simeq 10$\% given our $\simeq 10$\%
  uncertainty on $Y_{SZ}$ (see Figure~\ref{fig:ym}).
  
  Additionally, our results are consistent
  with $Y_{SZ}-M_{gas}$ scaling
  relations measured by other groups.
  For example, our results using model-independent
  $Y_{SZ}$ estimates from Bolocam and $M_{gas}$
  estimates from \emph{Chandra} within $r_{500}$
  agree well with the results in P10
  using $Y_{SZ}$ estimates from the
  SPT best-fit beta model and $M_{gas}$ estimates
  primarily from \emph{XMM-Newton}
  ($a = -5.73 \pm 0.43$ and $b = 2.12 \pm 0.45$
  within $r_{500}$ and 
  $a = -5.92 \pm 0.41$ and $b = 1.97 \pm 0.44$
  within $r_{2500}$).
  The results in \citet{bonamente08}, using
  beta model fits to 30~GHz OVRO/BIMA SZ data
  and \emph{Chandra} X-ray data within
  $r_{2500}$, also match our results quite
  well ($a = -5.22 \pm 1.77$ and $b = 1.41 \pm 0.13$).
  
\begin{deluxetable*}{cccccc} 
  \tablewidth{0pt}
  \tablecaption{Model-Independent $Y_{500}$ estimates}
  \tablehead{
    \colhead{} & \colhead{} &
    \colhead{JK noise} &
    \colhead{CMB/PS} & 
    \colhead{map DC signal} & 
    \colhead{flux cal.} \\
    \colhead{cluster} & 
    \colhead{$Y_{500}$} &
    \colhead{$\sigma_Y$} &
    \colhead{$\sigma_Y$} & 
    \colhead{$\sigma_Y$} & 
    \colhead{$\sigma_Y$}}
  \startdata
    Abell 697 & $59.2 \pm 6.3$ & 1.2 & 0.5 & 5.6 & 2.5 \\
    Abell 1835 & $47.5 \pm 4.9$ & 1.3 & 0.5 & 4.3 & 2.0 \\
    MS 0015.9+1609 & $23.3 \pm 2.2$ & 1.6 & 0.6 & 0.9 & 1.0 \\
    MS 0451.6-0305 & $24.7 \pm 2.0$ & 1.2 & 0.6 & 0.9 & 1.1 \\
    MS 1054.4-0321 & $8.8 \pm 1.0$ & 0.8 & 0.4 & 0.2 & 0.4 
  \enddata
  \tablecomments{
    Model independent estimates of $Y_{500}$, 
    in units of $10^{-11}$~ster, along with the 
    associated uncertainties.
    From left to right the columns give the value 
    and total uncertainty of our estimate of  
    $Y_{500}$, the uncertainty due to our 
    jackknife noise model (\emph{i.e.},
    excluding CMB and point sources), the
    uncertainty due to CMB and point sources, 
    the uncertainty in the DC signal level of the
    map (see Section~\ref{sec:tests_model}),
    and the uncertainty in our flux calibration.
    Note that the jackknife noise estimate is 
    dominated by large angular-scale residual
    atmospheric fluctuations; the noise
    on beam-size scales is negligible.}
  \label{tab:y_sz}
\end{deluxetable*}

\section{Tests for systematic errors}
  \label{sec:tests}

  We are primarily concerned with two types
  of systematic errors that may occur in 
  our analysis:
  those caused by our (minimal) use of 
  a cluster model, specifically an elliptical
  Nagai model, and those 
  caused by characteristics of our noise that
  are not properly accounted for in our analysis.
  We have run extensive tests to quantify the
  level of systematic error we can expect
  from each of these two sources.
  The details
  of these tests are 
  described below.
  In the end, we find that the amount of systematic
  error in our images is negligible,
  with the exception of the
  estimation of the DC signal offset in
  our images using the Nagai model.

  \subsection{Model dependence of results}
    \label{sec:tests_model}

    Since we compute the transfer function for 
    each cluster using the best-fit elliptical 
    Nagai model for that cluster,
    our deconvolved images necessarily
    have some model dependence.
    In order to quantify the amount of model
    dependence, we have computed transfer 
    functions for a range of elliptical Nagai
    models for one of the clusters in 
    our sample, MS 0451.6-0305.
    Relative to the best-fit model, we have
    varied the scale radius, $r_s$, the	
    ellipticity, $\epsilon$, the position
    angle, $\theta$, and the centroid location,
    $\delta$RA and $\delta$dec, by 
    increasing and decreasing each parameter
    individually by its $1 \sigma$ uncertainty\footnote{
      The power law slopes ($\mathcal{A}$, $\mathcal{B}$,
      and $\mathcal{C}$) were held fixed for all
      of our model fits. Due to the large 
      degeneracy between these values and 
      $r_s$, we have effectively included
      variations in the power law slopes
      by varying the value of $r_s$.}.
    We then deconvolved our processed map of MS 0451.6-0305
    using the transfer function computed from
    each model and subtracted the resulting map
    from the one produced using the transfer
    function for the best-fit model.
    In each case, the residual map was approximately
    white, with an RMS of 1.5, 0.6, and 0.5~\uK~for
    variations in $r_s$ and $\epsilon$, $\theta$, and
    $\delta$RA and $\delta$dec, respectively.
    Since the typical noise RMS of our deconvolved 
    maps is $\simeq 10$~\uK, the additional RMS
    introduced by our uncertainty in determining
    the model used for calculating a transfer
    function is quasi-negligible.
    Note that the best-fit elliptical Nagai model will not 
      provide an exact description of a real cluster.
      However, the elliptical Nagai model does
      provide an adequate description of four of the five
      clusters
      we have observed, indicating
      that the difference between the true cluster
      profile and the model profile is in general less than
      our noise.
      Therefore, the artifacts in our deconvolved map
      produced by using a model to describe the
      cluster will be smaller than the artifacts
      produced by our measurement uncertainty
      on the best-fit model.

    Additionally, we created a deconvolved map of 
    MS 0451.6-0305 using the transfer function for a point-like
    source.
    The resulting profile is significantly different
    from the profile obtained using the transfer
    function for the best-fit Nagai model,
    indicating that the naive calculation of a 
    transfer function using a point source 
    is inadequate.
    Compared to using the transfer function
    calculated from the best-fit elliptical
    Nagai model,
    the peak decrement is reduced by $\simeq 50$~\uK,
    while the magnitude of the 
    SZ signal at the edge of the map
    is increased by $\simeq 50$~\uK~(\emph{i.e.}, 
    the deconvolved cluster image from a point-source
    transfer function is systematically broader).

    As described in Section~\ref{sec:cluster_model}, we use the 
    best-fit elliptical Nagai model to determine the 
    DC signal level in our maps since its value
    is unconstrained by our data.
    The measurement uncertainty in the value of
    the DC signal is small, $< 1$~\uK, 
    but there is a large amount of uncertainty
    in the model at large radii.
    Based on the results from \citet{borgani04},
    N07, and \citet{piffaretti08} presented in
    \cite{arnaud09}, the RMS 
    scatter
    in the pressure profiles
    from cluster to cluster in simulations is
    $\lesssim 25$\%.
    Therefore, we include an additional 25\%
    systematic uncertainty on the DC signal
    that we add to the deconvolved map.
    Specifically, we estimate that the model
    uncertainty in the DC signal level
    of our map is 25\% of the signal
    level at the edge of the map.
    
  \subsection{Noise characteristics}
    \label{sec:tests_noise}

\begin{figure*}
  \plottwo{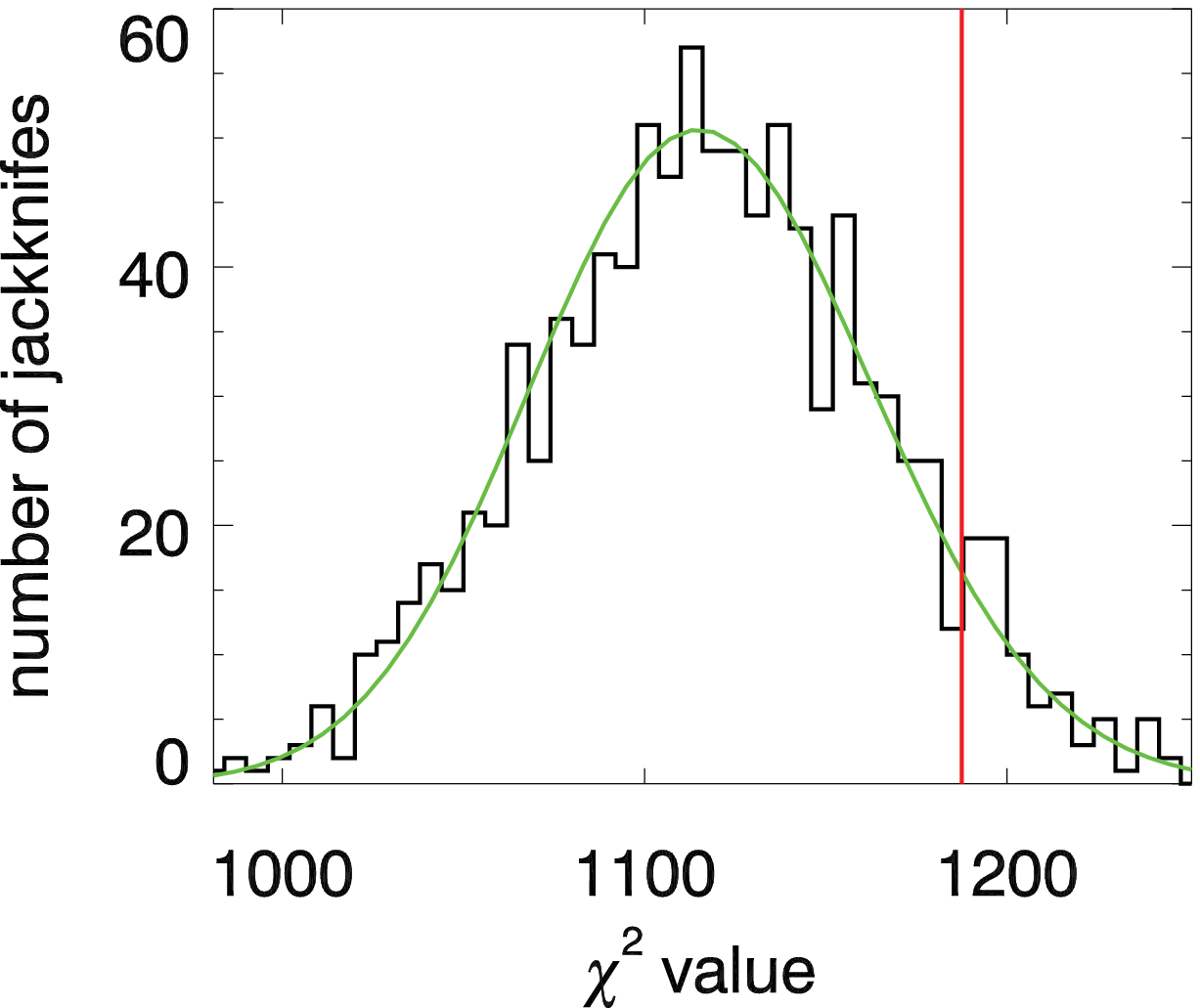}{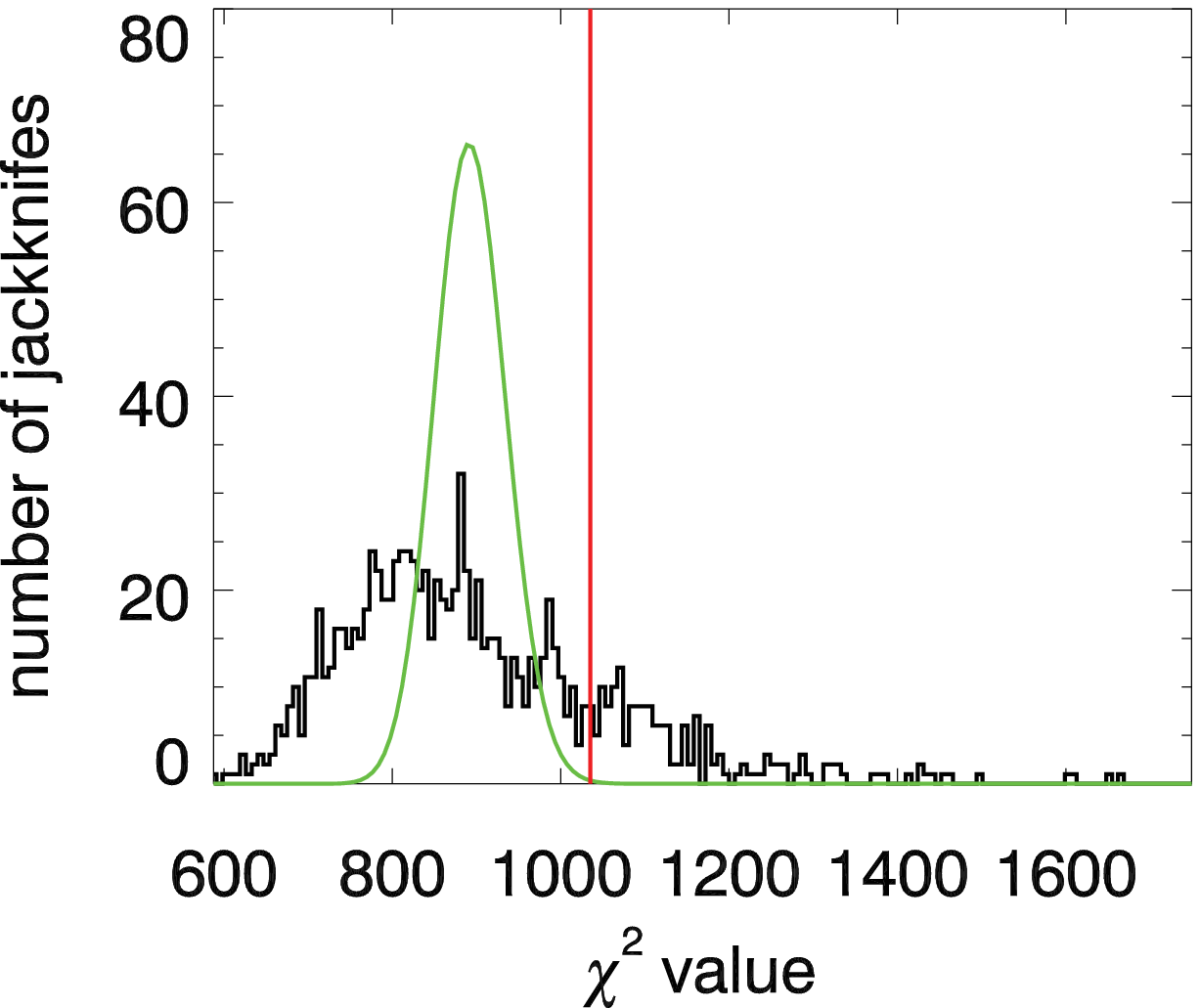}
  \caption{Histograms of the $\chi^2$ value for 1000
    separate noise realizations for MS 0451.6-0305, 
    overplotted in green with the predicted distribution
    assuming the noise covariance matrix is 
    diagonal (\emph{i.e.}, there is no correlated noise).
    For each noise realization the best-fit elliptical
    Nagai model profile for MS 0451.6-0305 is added to
    the noise realization, 
    an elliptical Nagai model is fit to this
    model-cluster-plus-noise realization,
    and the value of 
    $\chi^2$ is computed based on the assumption
    that the noise covariance matrix is diagonal.
    The vertical red line shows the value of 
    $\chi^2$ for the actual data for MS 0451.6-0305.
    The left histogram shows the processed data,
    and the right histogram shows the data for
    the deconvolved image.
    The predicted and actual $\chi^2$ distributions
    for the processed data overlap, indicating that
    there are minimal correlations between map pixels.
    However, the actual $\chi^2$ distribution for the
    deconvolved image data is much broader than the 
    predicted distribution, indicating that
    there are significant noise correlations between
    map pixels in the deconvolved images.}
  \label{fig:chisq}
  \vspace{20pt}
\end{figure*}

    As mentioned in Section~\ref{sec:reduction}, we make the 
    approximation that the noise
    covariance matrix is diagonal in our processed maps (\emph{i.e.},
    there are no noise correlations between pixels).
    Although fluctuations in both the atmospheric emission
    and the CMB are correlated over many pixels, the 
    high-pass filter we apply to our data timestreams
    eliminates these correlations within our ability
    to measure them.
    To test for noise correlations, we added
    processed cluster images of the best-fit
    models to
    the 1000 noise realizations for each cluster
    (\emph{i.e.}, for a given cluster we 
    separately added each of the four best-fit
    cluster models from Tables~\ref{tab:nagai_model}
    and \ref{tab:beta_model} to each of the 1000
    noise realizations).
    We then fit a model of the same type
    (\emph{e.g.}, if we added the best-fit
      elliptical Nagai model to the noise realization,
      then we fit an elliptical Nagai model
      to the cluster-model-plus-noise map)
    to each of these model-cluster-plus-noise maps
    and examined the distribution
    of $\chi^2$ values for these fits.
    For all five clusters,
    we found that the distribution of $\chi^2$ values
    matched the predicted $\chi^2$ distribution
    obtained from the diagonal noise covariance
    matrix\footnote{
      We have fit for $N_{params}$ free parameters
      in our fits to both the actual data and the 
      noise realizations,
      with $N_{params} = 5$ or 7 for the 
      spherical or elliptical fits.
      Therefore, the predicted $\chi^2$
      distribution is for $N_{pix} - N_{params}$
      DOF, where $N_{pix}$ is the number of map
      pixels.}.
    As an example, the fit quality of 
    our measured $\chi^2$
    distribution for the elliptical-Nagai-model-plus-noise
    realizations to the predicted $\chi^2$ distribution,
    quantified by a PTE,
    is 0.58, 0.02, 0.08, 0.66, and 0.21 for
    Abell 697, Abell 1835, MS 0015.9+1609, 
    MS 0451.6-0305, and MS 1054.4+321.
    (see Figure~\ref{fig:chisq}).
    Therefore, we conclude that there is a 
    negligible amount of correlated noise
    in our processed images and our assumption
    that the noise covariance matrix is diagonal
    is valid.
    Furthermore, we have estimated all of 
    our parameter uncertainties ($p_0, 
    r_s$,
    etc.) directly from the distribution
    of values calculated from our noise realizations.
    Consequently, any failure of our assumption
    that the covariance matrix is diagonal
    for the processed images
    will only affect the pixel-weighting
    and $\chi^2$ values for our model fits.

    However, when we perform the same test
    of fitting a model to our model-cluster-plus-noise
    maps
    using noise realizations that have been
    deconvolved with our transfer function,
    the result is significantly different.
    The distribution of $\chi^2$ values
    calculated from our deconvolved noise
    realizations is significantly broader
    than the predicted distribution based
    on uncorrelated noise
    (see Figure~\ref{fig:chisq}).
    This result is not surprising since the 
    deconvolution enhances the large-scale
    signals in the images, including
    residual atmospheric noise and 
    CMB fluctuations.
    Since the noise in the deconvolved images
    is significantly spatially correlated, our assumption
    that the noise covariance matrix is 
    diagonal fails.
    Therefore, we estimate the uncertainties
    for the deconvolved images using the 
    spread in values for the 
    noise realizations rather than
    from the diagonal elements of the noise covariance
    matrix 
    (\emph{e.g.}, the uncertainties in the radial
    profiles are determined from the RMS spread in the 
    radial profiles of the noise realizations).
    This is the same technique used by
    \citet{nord09} and \citet{basu10} to 
    analyze APEX-SZ data.

    The model
    fits to the model-cluster-plus-noise realizations
    also provide us with estimates of the uncertainties
    and biases associated with our model-parameter
    fitting.
    Specifically, we obtain 1000 best-fit values
    for each parameter;
    the standard deviation of these values
    then gives the uncertainty on our
    estimate of that parameter in our
    actual data map.
    These uncertainties are given in 
    Tables~\ref{tab:nagai_model} and 
    \ref{tab:beta_model}.
    In addition,
    if our parameter estimation algorithm is 
    free from biases, then we should,
    on average, recover the parameters of the
    input model that was added to the noise
    realizations.
    In practice, we find a small, but measurable,
    bias in our estimates of the pressure
    normalization and scale radius in 
    our fits;
    the bias is typically $\lesssim 10$\%
    of the uncertainty on each paramter.
    We find no measurable bias in our estimates
    of the other fit parameters.

    Additionally, we have used our signal-free SDS1
    maps to further verify our model-fitting proceedure and 
    to search for any components of the noise that
    have not been included in our noise estimate.
    First, we 
    inserted model clusters into the SDS1 data
    timestreams based on the
    best-fit elliptical Nagai profile for
    each of the five clusters in our sample.
    These data were then processed and 
    an elliptical Nagai model was fit to
    each resulting image.
    In each fit, there are 6 free parameters
    ($p_0$, $r_s$, $\epsilon$, $\theta$, 
    $\delta$RA, and $\delta$dec), giving
    us a total of 30 fit parameters for the 
    5 model clusters.
    Of these 30 fit parameters, 17 (57\%) are
    within $1\sigma$ of the input value,
    26 (87\%) are within $2\sigma$ of the 
    input value, and all 30 are within
    $3\sigma$ of the input value;
    these results
    indicate that our model fitting
    and parameter error estimation
    are working properly.
    Additionally, we obtain a reasonable
    goodness of fit for the models using
    these best-fit parameters,
    quantified by a PTE $\simeq 0.8$,
    providing further evidence
    that there is no significant noise
    in the data that has not been
    included in our noise estimate.
    Note that since the five cluster model profiles are
      fairly similar, we obtain
      comparable PTEs for all five profiles.

\section{Summary}

  We have presented the first results from our program to
  image the SZ effect in galaxy clusters with Bolocam.
  These images have a beam-smoothed RMS of $\simeq 10$~\uK,
  and a resolution of 58~arcsec. 
  Given this noise level, we are able to measure
  SZ signal in radial profiles to approximately the edge of our
  maps, which corresponds to 6-7~arcmin
  or $1-2$ times $r_{500}$.
  In order to subtract noise from atmospheric fluctuations,
  we effectively high-pass filter our cluster
  images.
  However, we are able to deconvolve the effects
  of this filter with biases that are negligible
  compared to our noise level,
  other than our recovery of the DC signal level.
  In fitting our images to spherical and elliptical
  beta and Nagai models, we find no
  preference between the beta and Nagai
  models due to the degeneracy between
  these models over the angular range
  to which our data are sensitive,
  but our data do show a definitive
  preference for elliptical models
  over spherical models.
  The weighted mean ellipticity of the five
  clusters is $\epsilon = 0.27 \pm 0.03$, 
  consistent with results from X-ray data.
  Additionally, the best-fit model parameters we determine from
  our data are consistent with those found
  from previous X-ray and SZ measurements.
  We have also obtained model-independent estimates
  of $Y_{SZ}$, and we find scaling relations
  between $Y_{SZ}$ and cluster mass that
  are consistent with self-similar predictions,
  with a scatter that is consistent with
  expectations for a $\simeq 10$\% intrinsic scatter.

\section{Acknowledgments}

We acknowledge the assistance of: 
Nicole Czakon, Tom Downes, and Seth Siegel,
who have provided numerous comments and suggestions
about the data analysis;
the Bolocam instrument team:
P.~A.~R. Ade, J.~E. Aguirre, J.~J. Bock, S.~F. Edgington,
J. Glenn, A. Goldin, S.~R. Golwala, 
D. Haig, A.~E. Lange, G.~T. Laurent,
P.~D. Mauskopf, H.~T. Nguyen, P. Rossinot, and J. Sayers;
Matt Ferry, who helped collect the data for
Abell 1835 and MS 1054.4-0321;
the day crew and Hilo
staff of the Caltech Submillimeter Observatory, who provided
invaluable assistance during commissioning and data-taking for this
survey data set; 
Daisuke Nagai for useful discussions about cluster modeling;
Tony Mroczkowski for providing the best-fit parameters
from his group's analysis of Abell 1835,
useful comments about cluster modeling,
and pointing out a typo in our original arxiv posting;
the referee for several useful comments and suggestions;
and Kathy Deniston, Barbara Wertz, and Diana Bisel, who provided effective
administrative support at Caltech and in Hilo.  Bolocam was constructed and
commissioned using funds from NSF/AST-9618798, NSF/AST-0098737,
NSF/AST-9980846, NSF/AST-0229008, and NSF/AST-0206158.  JS 
was partially supported by a
NASA Graduate Student Research Fellowship and a NASA
Postdoctoral Program fellowship,
SA and EP were partially supported by NSF grant AST-0649899,
SA was partially supported by the USC WiSE postdoctoral
fellowship and travel grants, and
EP was partially supported by NASA grant NNXO7AH59G
and JPL-Planck subcontract 1290790.

{\it Facilities:} \facility{CSO}.

\appendix
\section{}
  \label{sec:appendix}

  This appendix includes images and radial profiles of the processed and
  deconvolved maps, an image of the 
  processed residual map after subtracting
  the best-fit elliptical Nagai model,
  and an image of one of the 1000 noise estimates generated
  via jackknife realizations of the data and a model
  for the astronomical noise.
  We have smoothed all of the images using a Gaussian
  beam with a FWHM of 58~arcsec.
  A white dot representing the FWHM of the effective
  PSF for these beam-smoothed images is given in the 
  lower left of each image.
  The solid white contour lines in the images represent a 
  S/N of $-2,-4,..$, and the dashed white contour lines 
  represent a S/N of $+2,+4,...$.
  The deconvolved images contain a significant amount 
  of noise that is correlated over large angular scales,
  along with a model-dependent DC signal offset,
  and we therefore do not display noise contours
  on the deconvolved images.
  The error bars on the radial profiles are estimated
  from the spread in radial profiles computed
  from our noise realizations, and therefore
  do include all of the large-angular-scale noise correlations
  (although they do not include the
  uncertainty in the DC signal level of the image).
  Note that the radial profile bins for the deconvolved
  images are correlated due to the large-angular-scale
  noise present in those images.

  \setcounter{figure}{0}
  \renewcommand{\thefigure}{A\arabic{figure}}

  \clearpage
  \begin{figure*}
    \centering
    \includegraphics[height=.25\textheight]{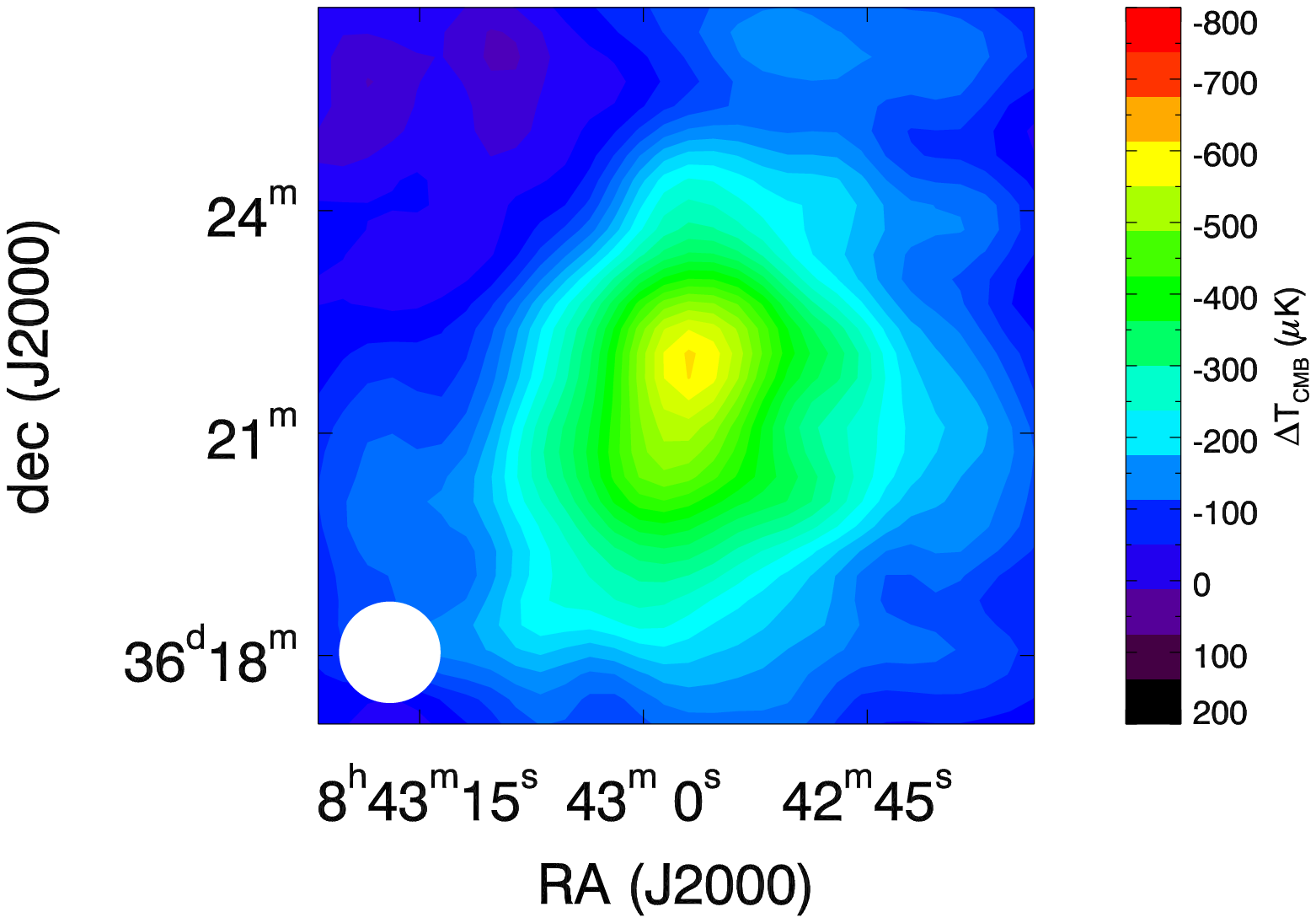} \hspace{.025\textwidth}
    \includegraphics[height=.25\textheight]{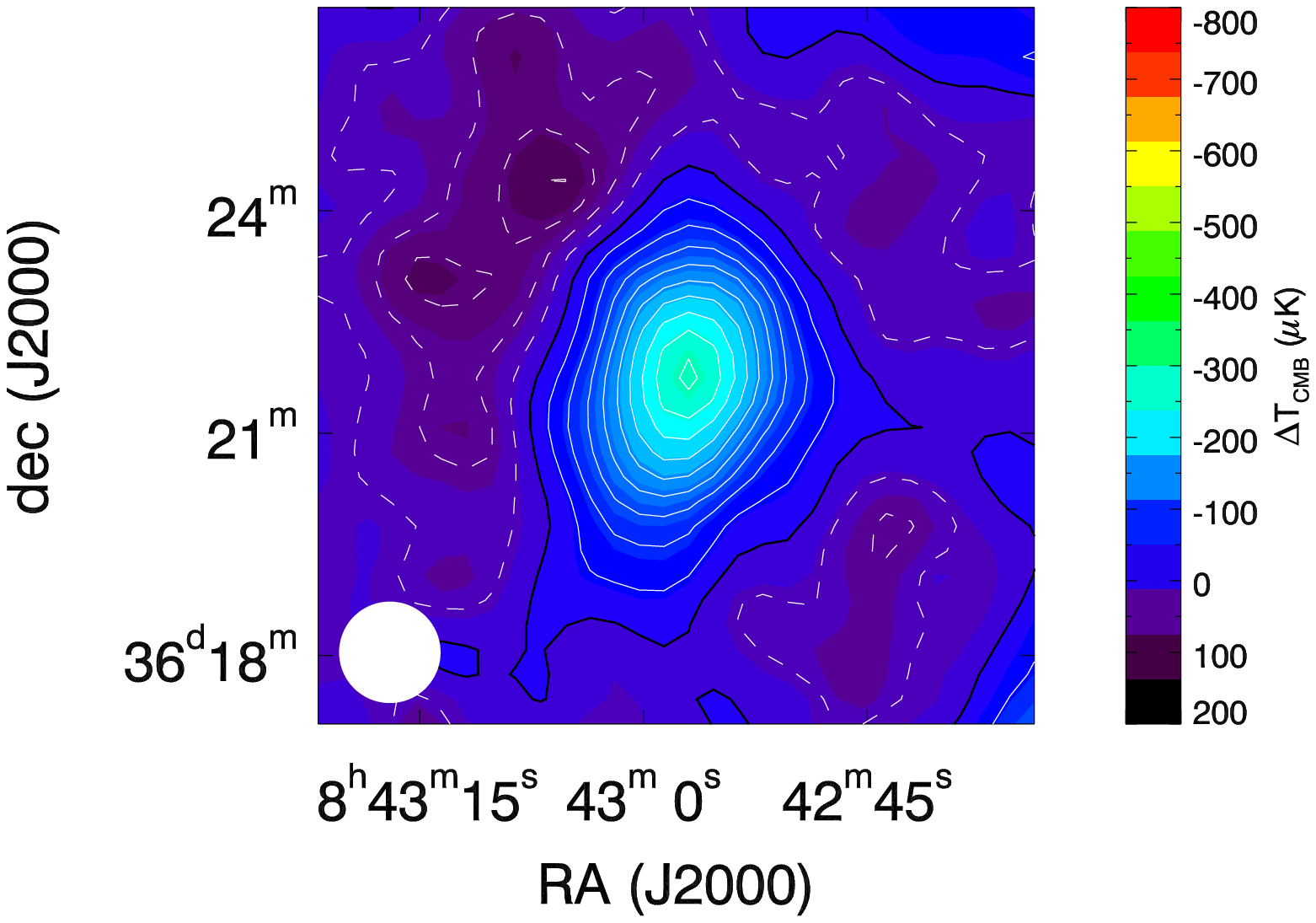} \hspace{.025\textwidth}

    \vspace{.02\textheight}
    \centering
    \includegraphics[height=.25\textheight]{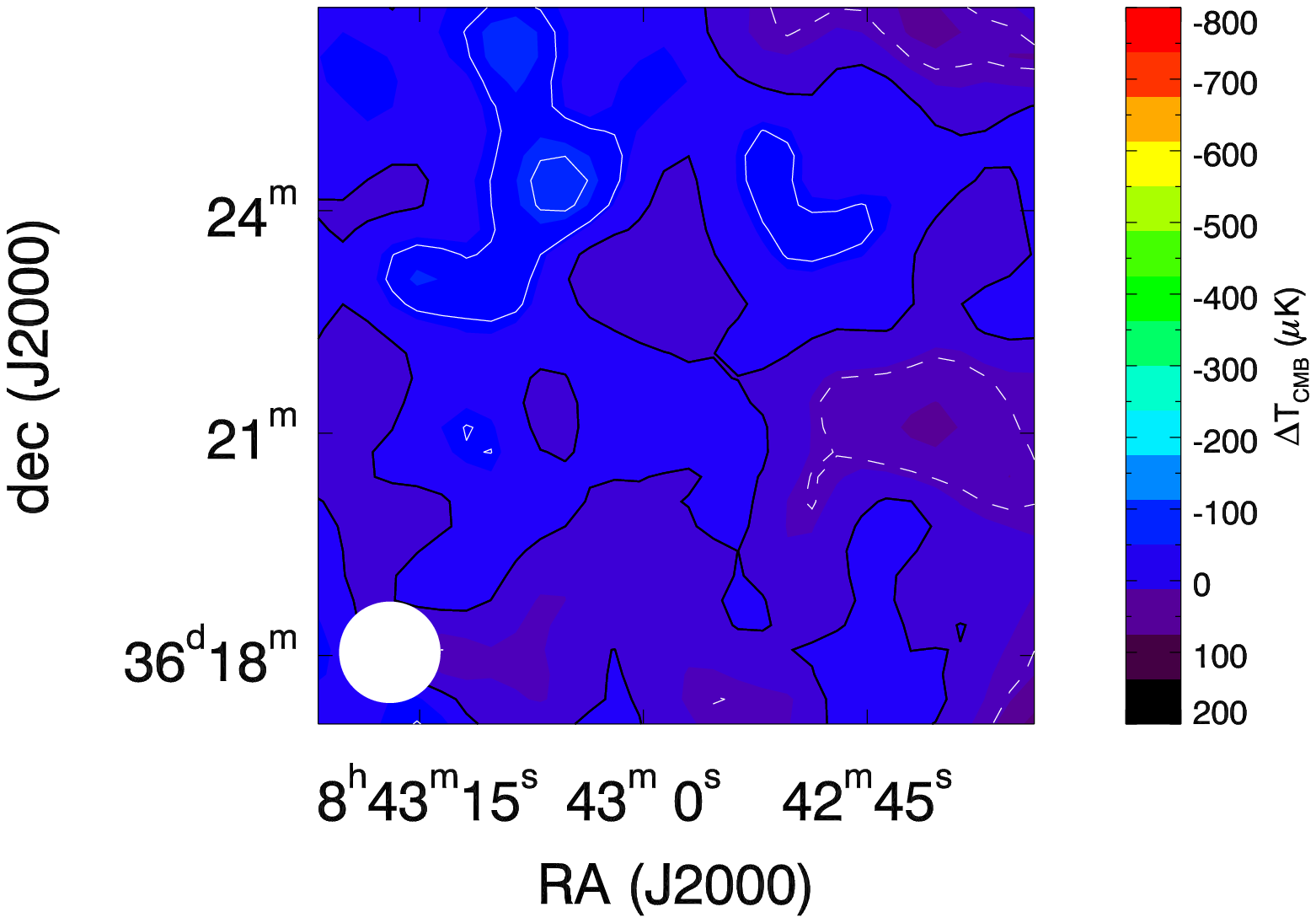} \hspace{.025\textwidth}
    \includegraphics[height=.25\textheight]{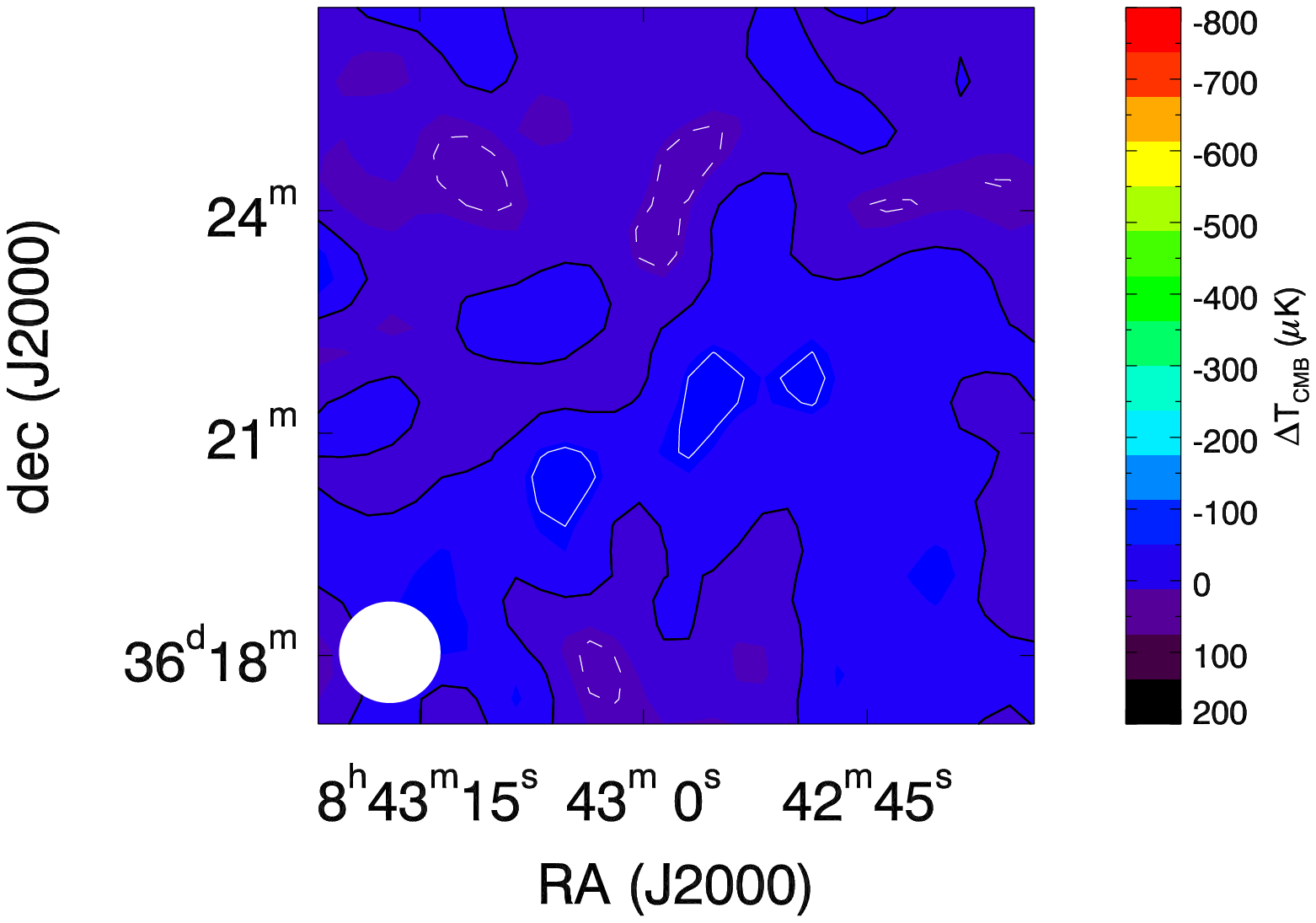} \hspace{.025\textwidth}

    \vspace{.02\textheight}
    \centering
    \includegraphics[height=.35\textheight]{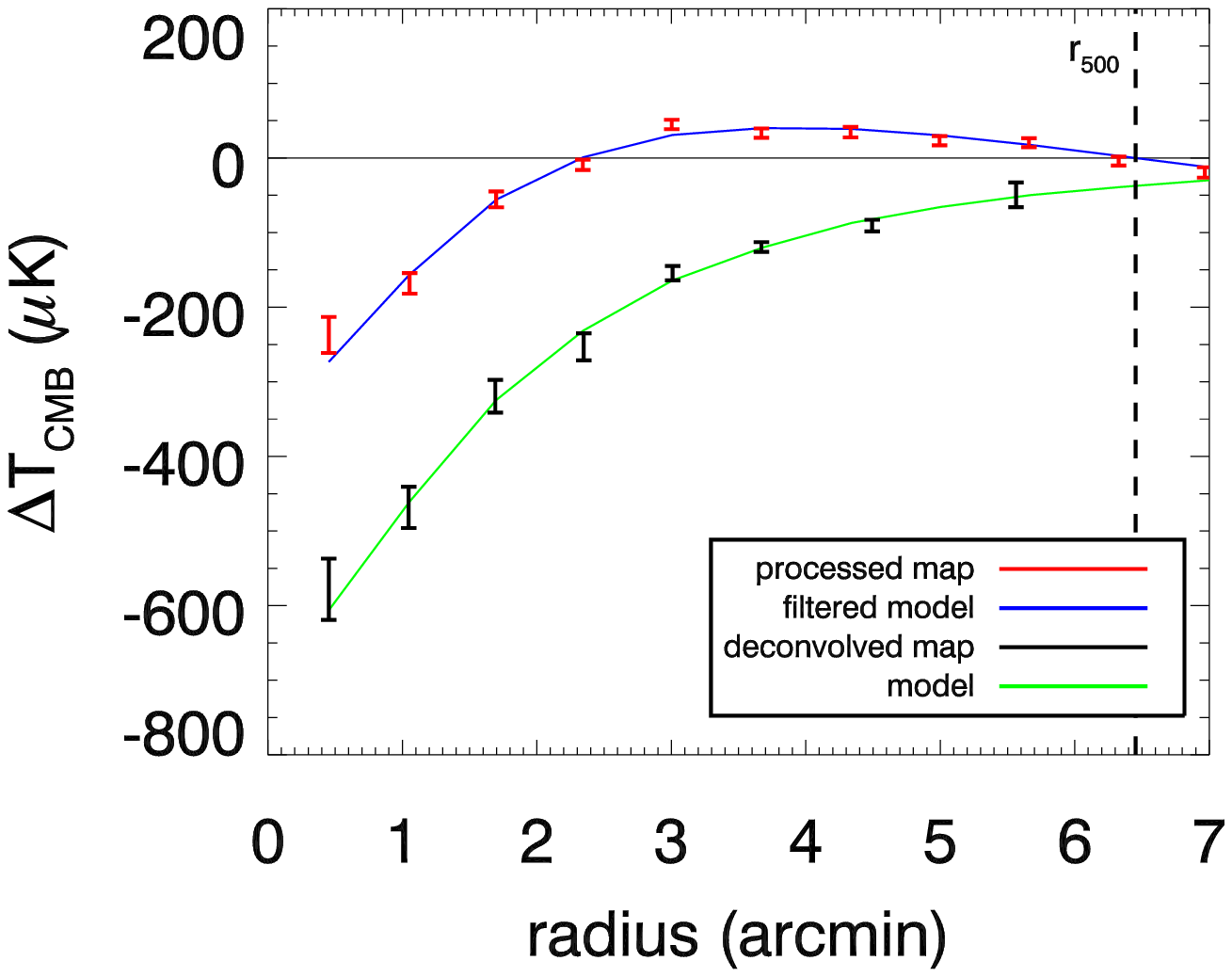}
    \caption{Abell 697; from left to right and top to bottom we show
      the deconvovled image of the cluster, 
      the processed image of the cluster,
      the residual map between the processed image of the 
      cluster and the best-fit elliptical Nagai model, 
      one of the 1000 noise realizations for the processed data,
      and a binned radial profile.
      The contour lines represent a S/N of $2,4,..$.}
    \label{fig:a697}
  \end{figure*}

  \clearpage
  \begin{figure*}
    \centering
    \includegraphics[height=.25\textheight]{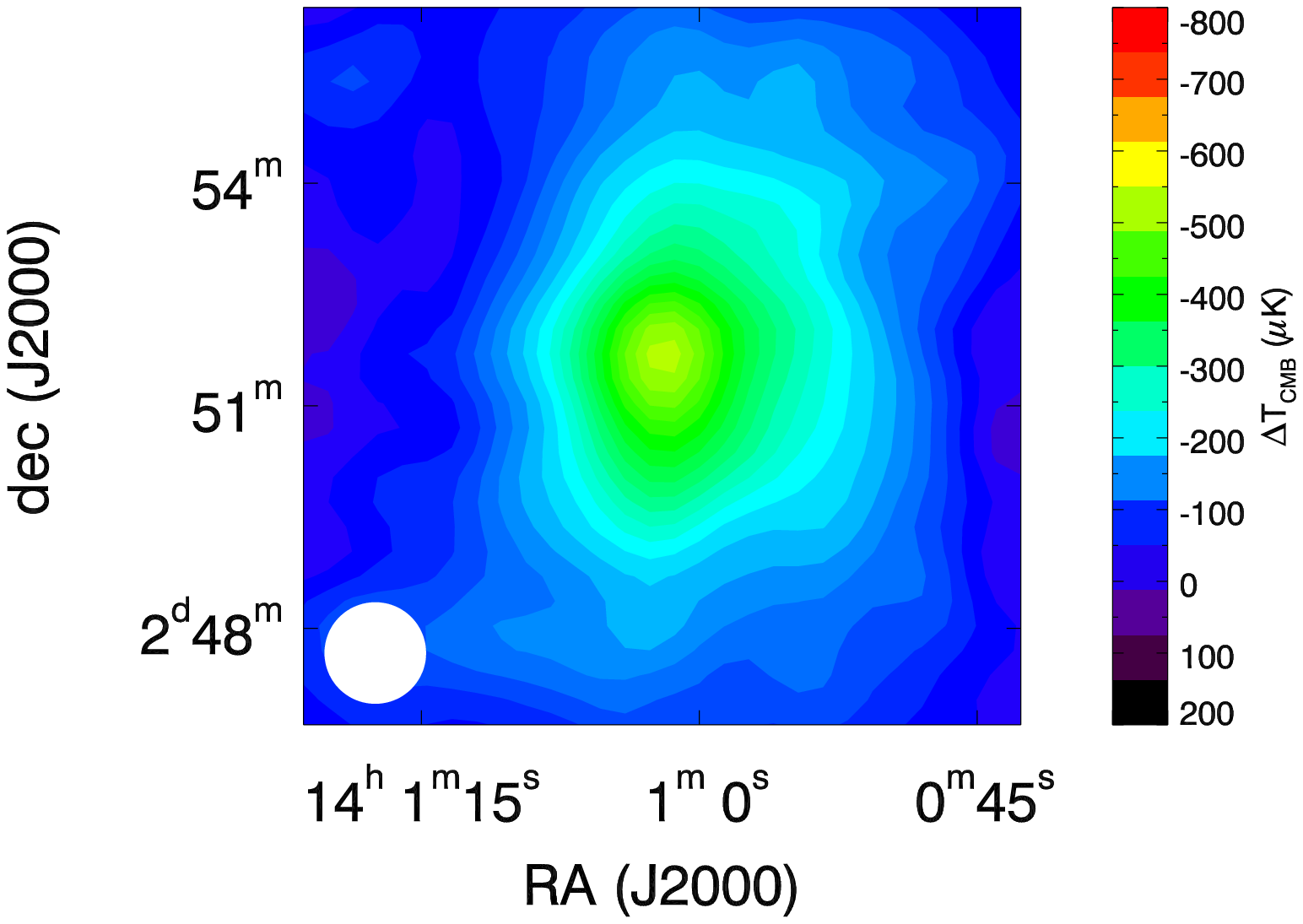} \hspace{.025\textwidth}
    \includegraphics[height=.25\textheight]{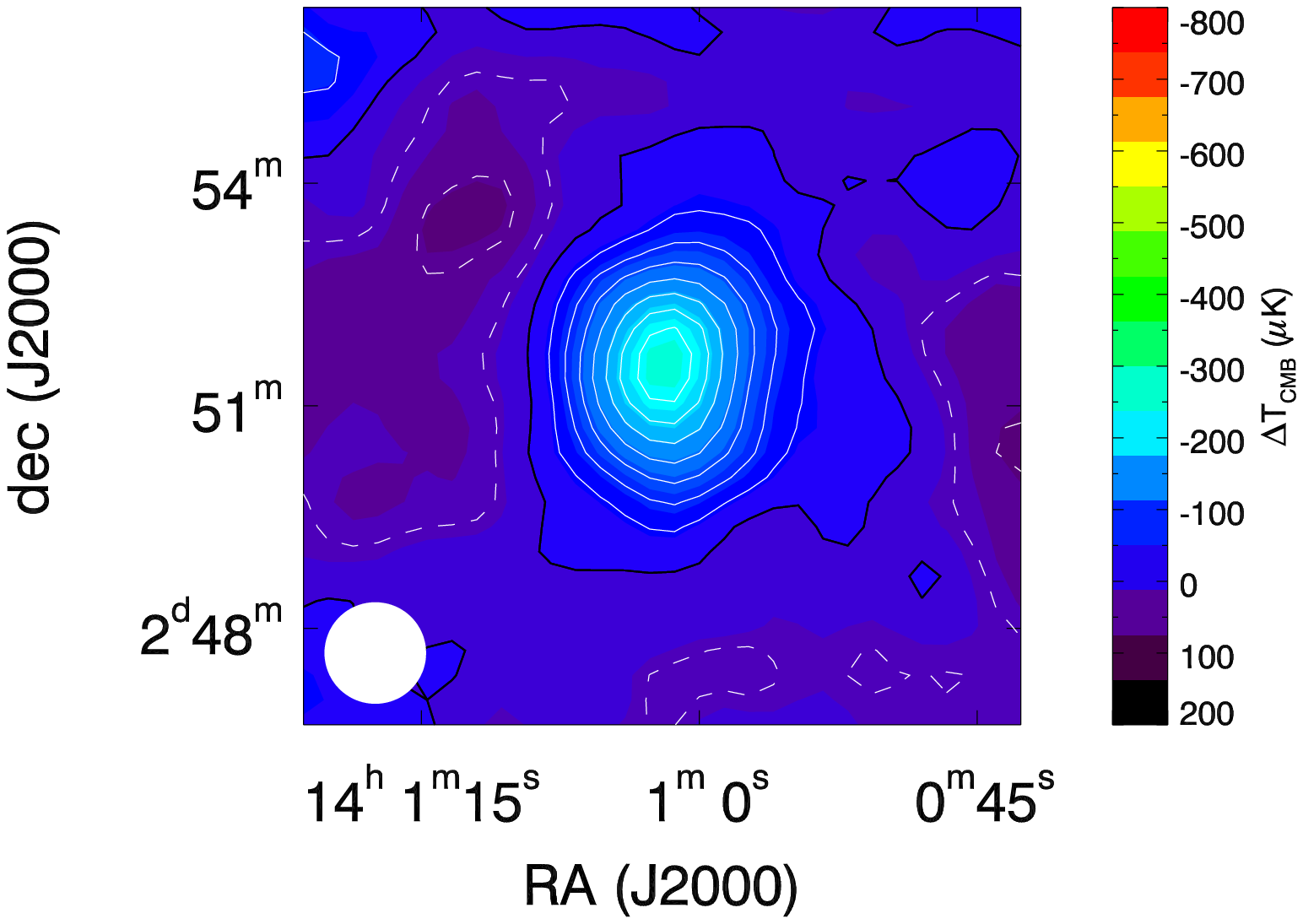} \hspace{.025\textwidth}

    \vspace{.02\textheight}
    \centering
    \includegraphics[height=.25\textheight]{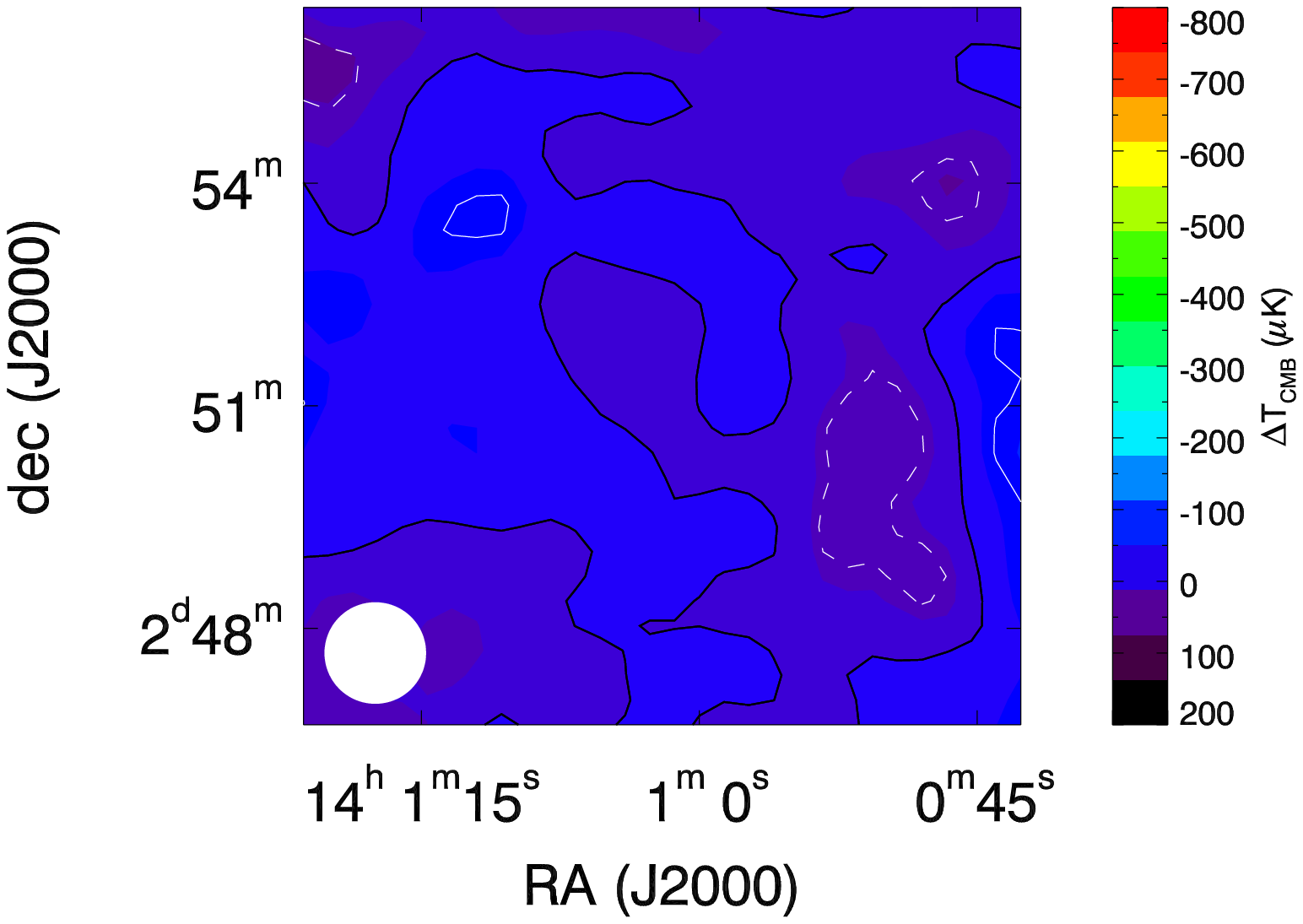} \hspace{.025\textwidth}
    \includegraphics[height=.25\textheight]{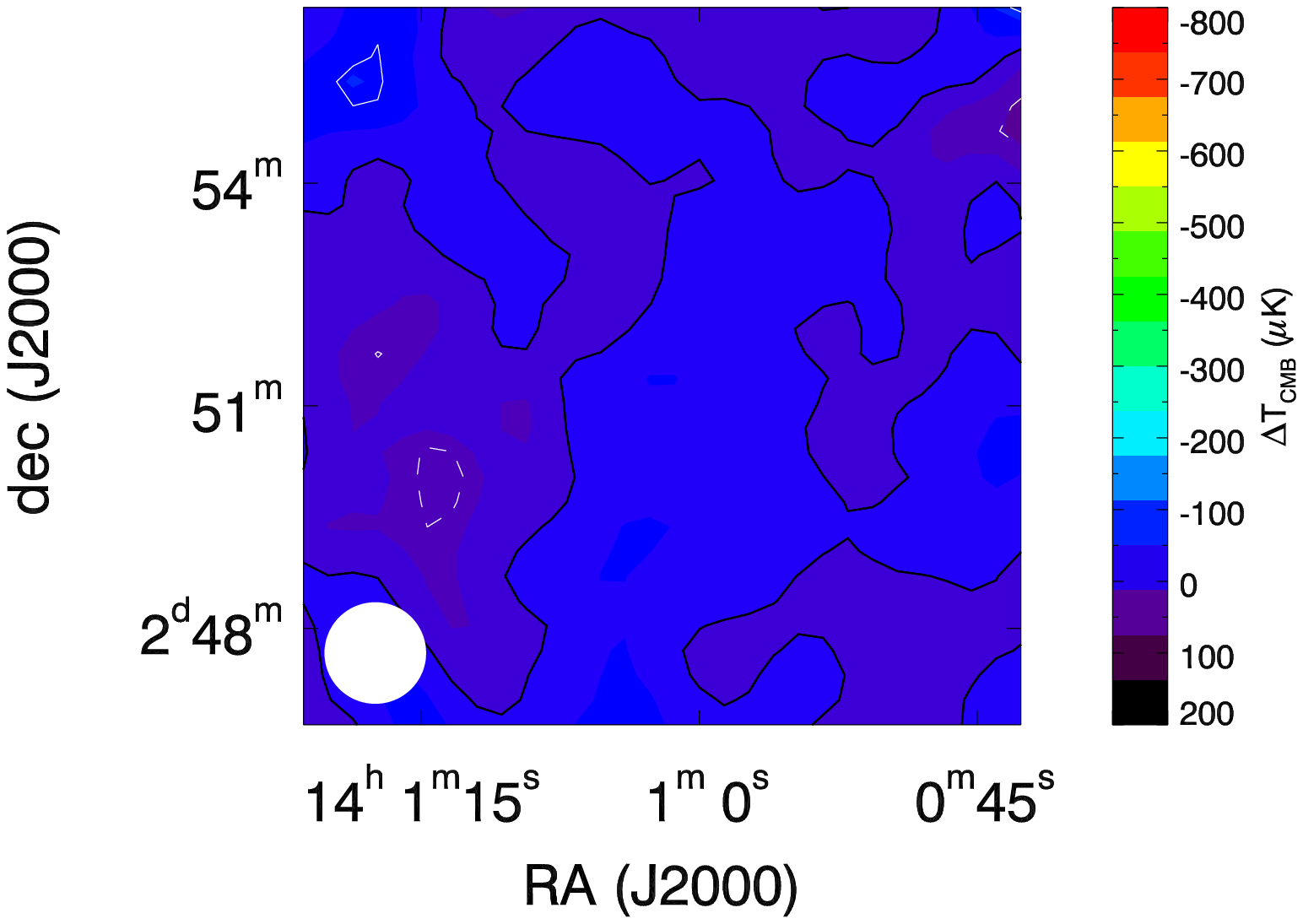} \hspace{.025\textwidth}

    \vspace{.02\textheight}
    \centering
    \includegraphics[height=.35\textheight]{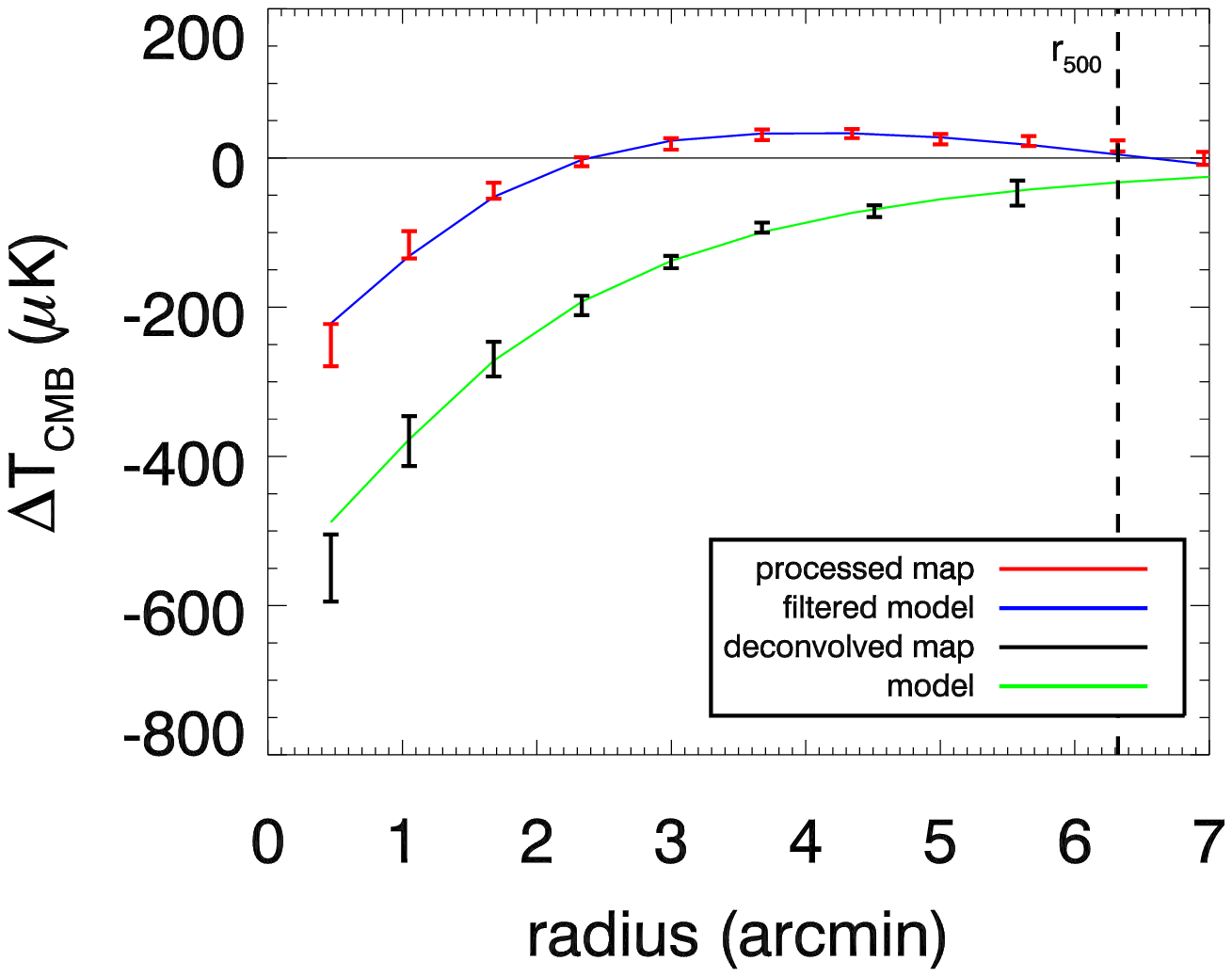}
    \caption{Abell 1835; from left to right and top to bottom we show
      the deconvovled image of the cluster, 
      the processed image of the cluster,
      the residual map between the processed image of the 
      cluster and the best-fit elliptical Nagai model, 
      one of the 1000 noise realizations for the processed data,
      and a binned radial profile.
      The contour lines represent a S/N of $2,4,..$.}
    \label{fig:a1835}
  \end{figure*}

  \clearpage
  \begin{figure*}
    \centering
    \includegraphics[height=.25\textheight]{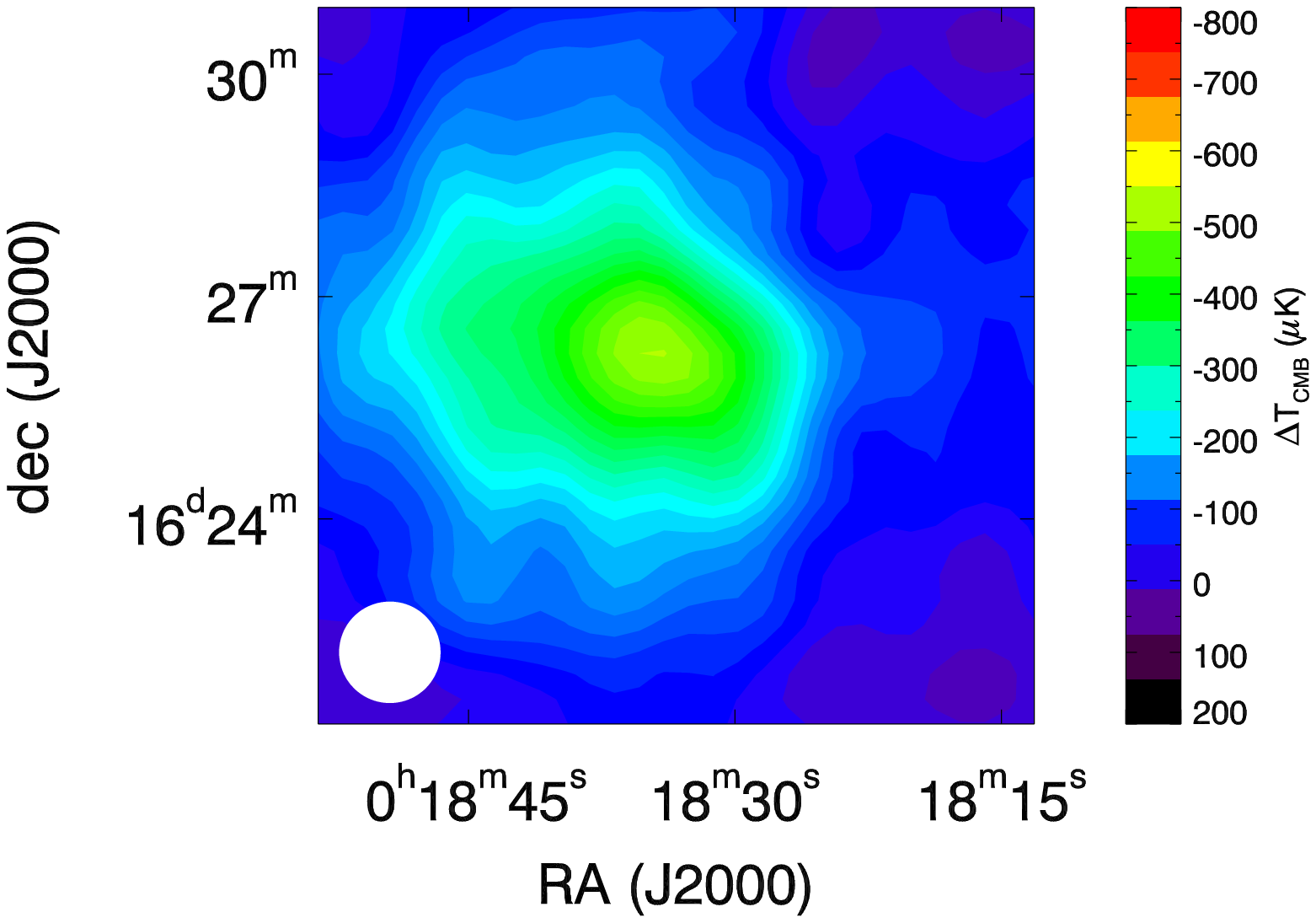} \hspace{.025\textwidth}
    \includegraphics[height=.25\textheight]{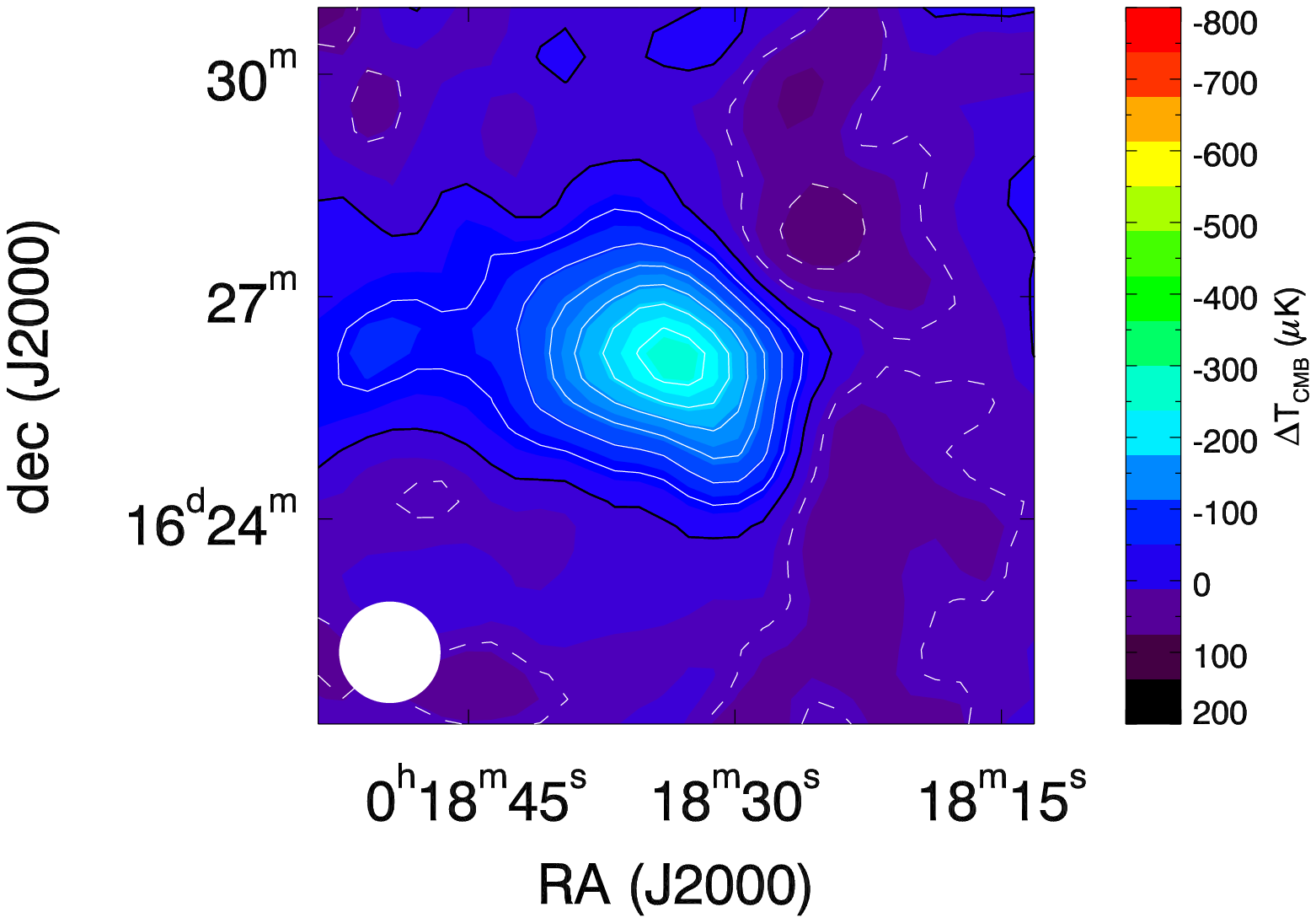} \hspace{.025\textwidth}

    \vspace{.02\textheight}
    \centering
    \includegraphics[height=.25\textheight]{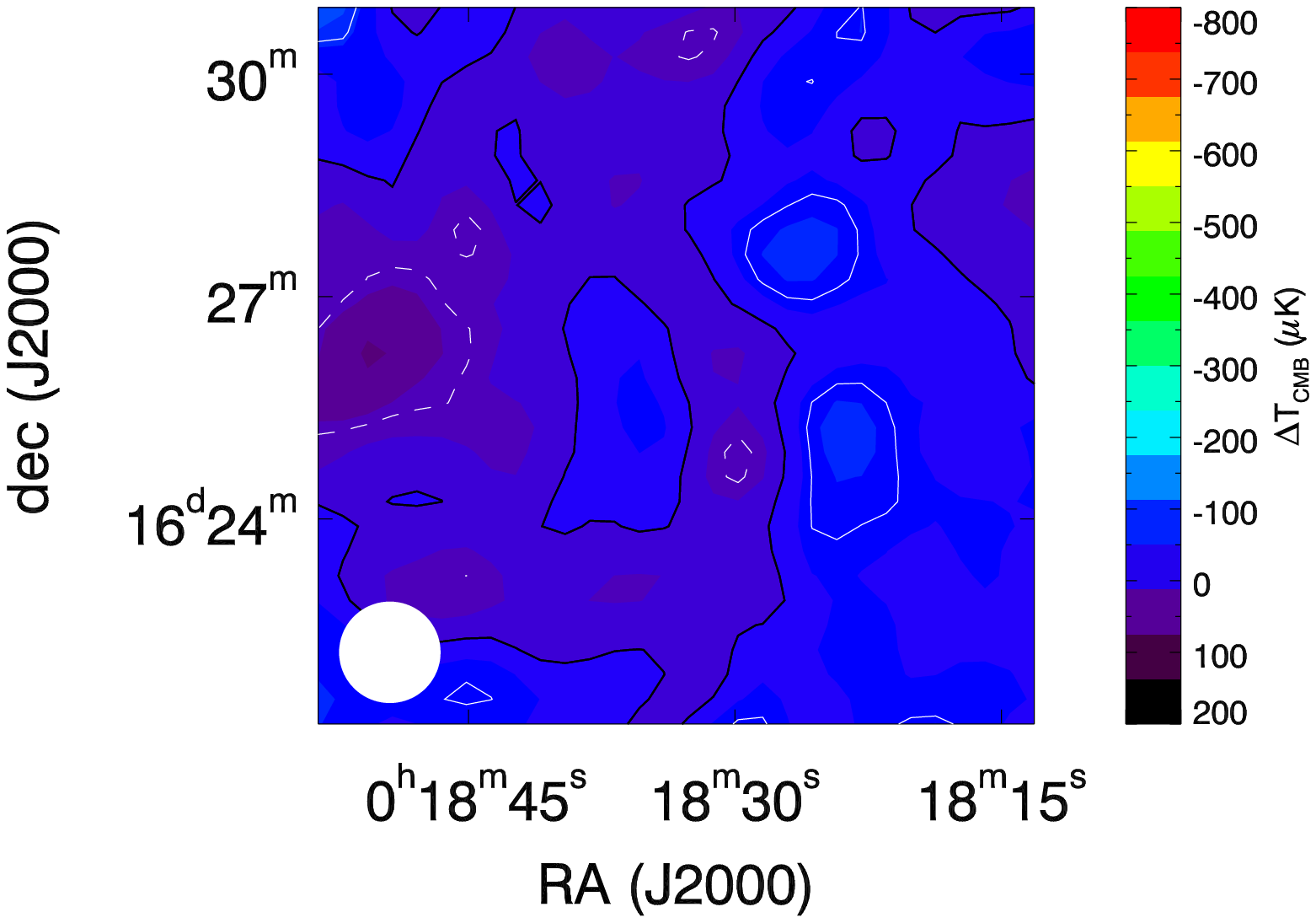} \hspace{.025\textwidth}
    \includegraphics[height=.25\textheight]{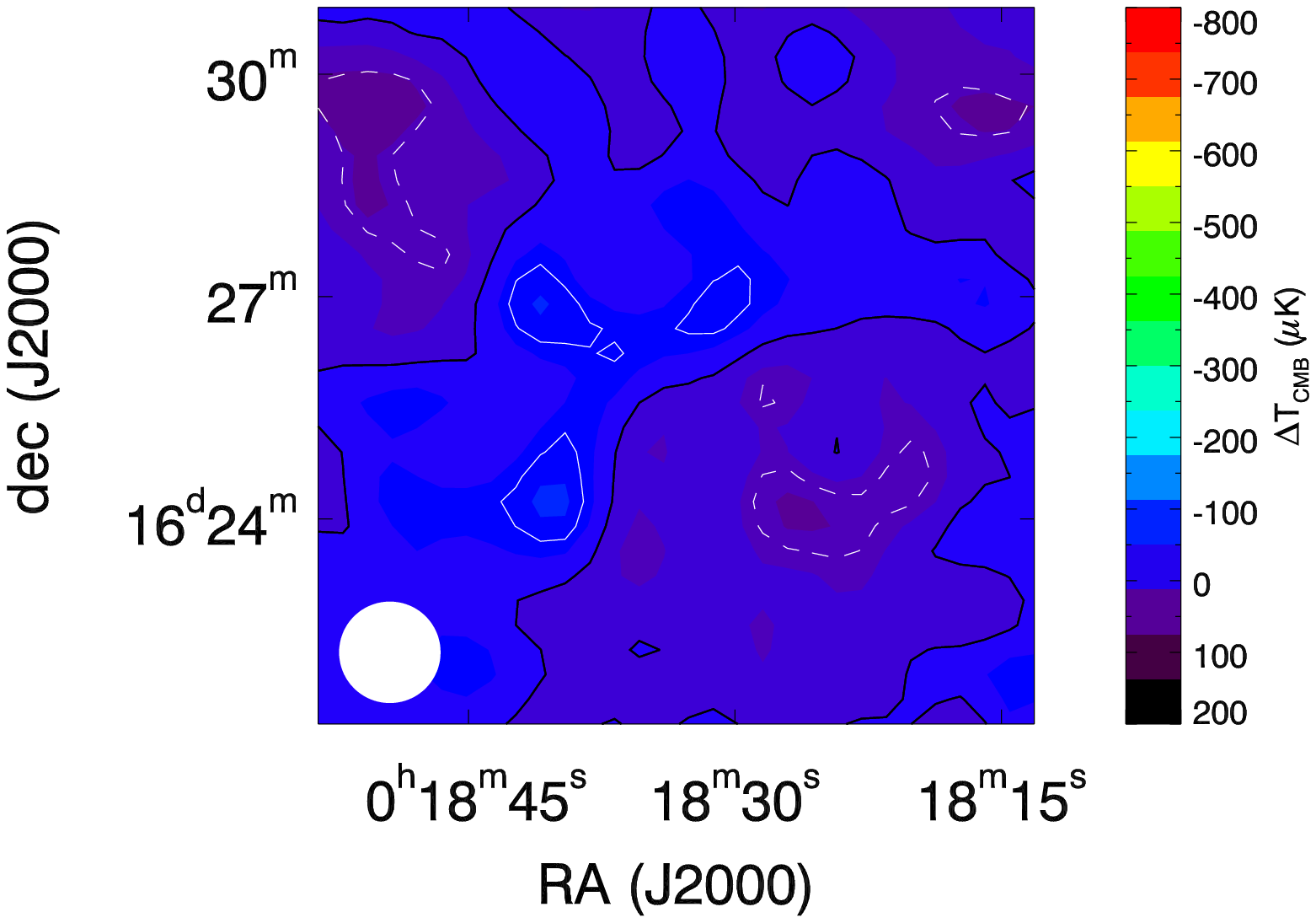} \hspace{.025\textwidth}

    \vspace{.02\textheight}
    \centering
    \includegraphics[height=.35\textheight]{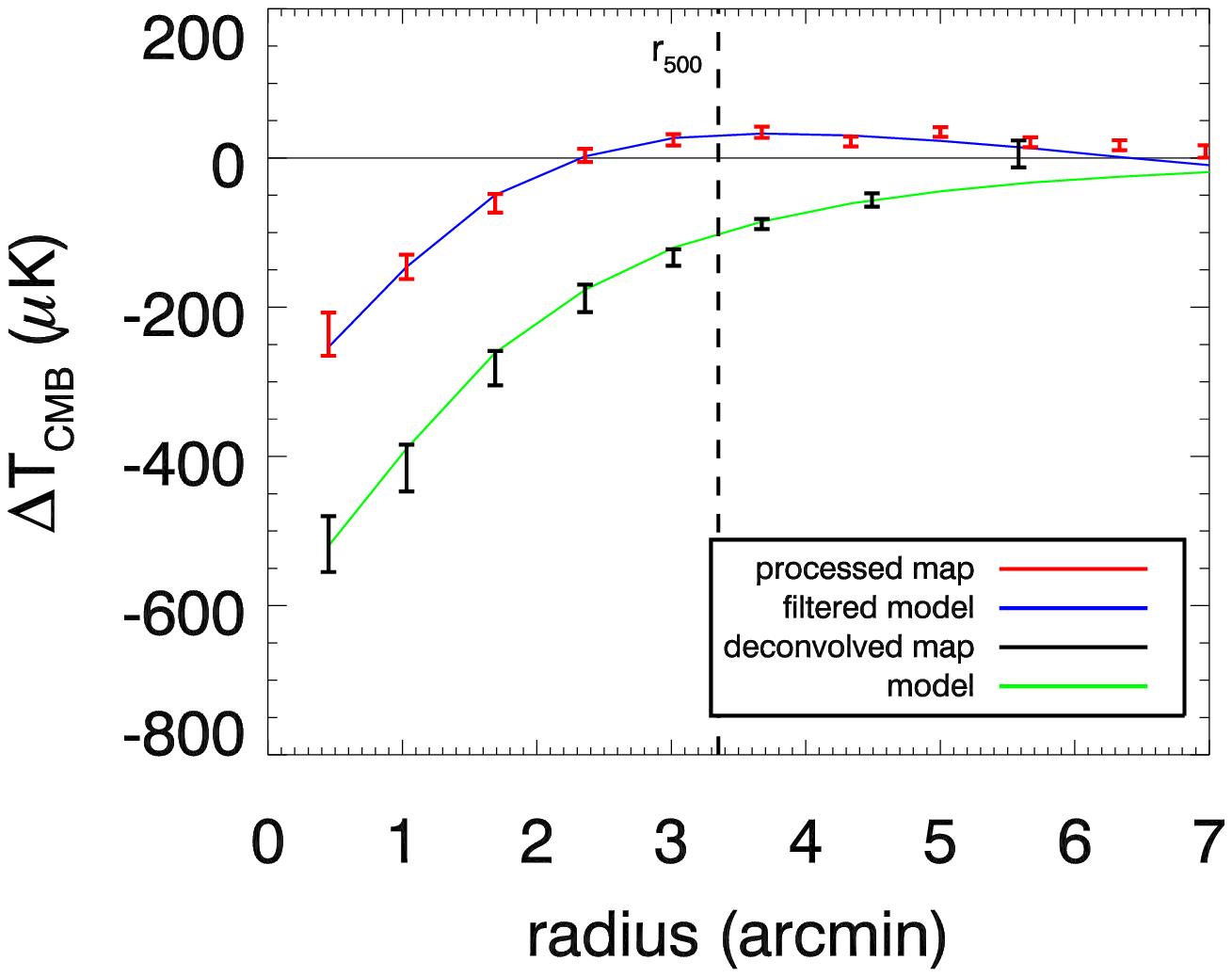}
    \caption{MS 0015.9+1609; from left to right and top to bottom we show
      the deconvovled image of the cluster, 
      the processed image of the cluster,
      the residual map between the processed image of the 
      cluster and the best-fit elliptical Nagai model, 
      one of the 1000 noise realizations for the processed data,
      and a binned radial profile.
      The contour lines represent a S/N of $2,4,..$.}
    \label{fig:ms0016}
  \end{figure*}

  \clearpage
  \begin{figure*}
    \centering
    \includegraphics[height=.25\textheight]{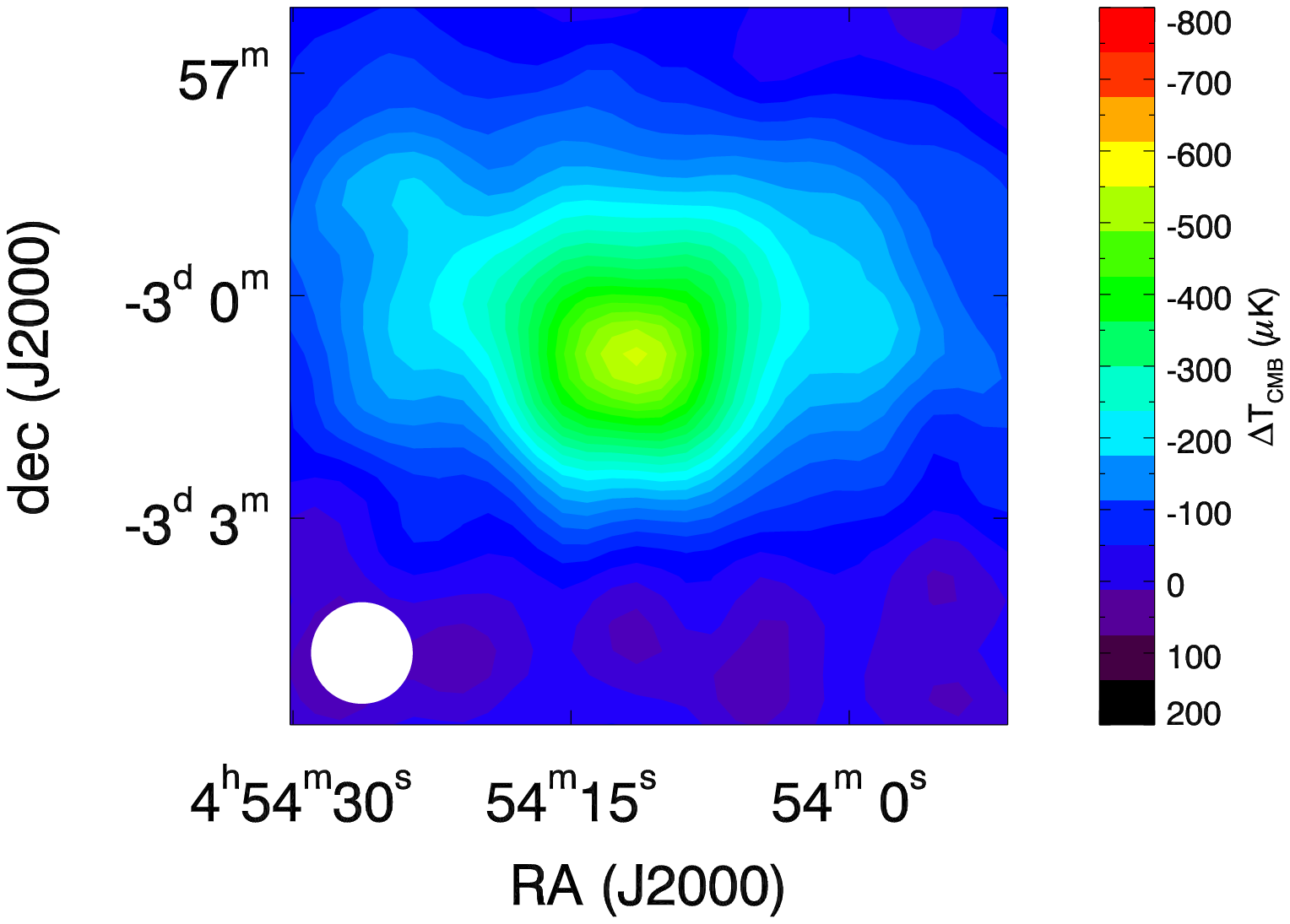} \hspace{.025\textwidth}
    \includegraphics[height=.25\textheight]{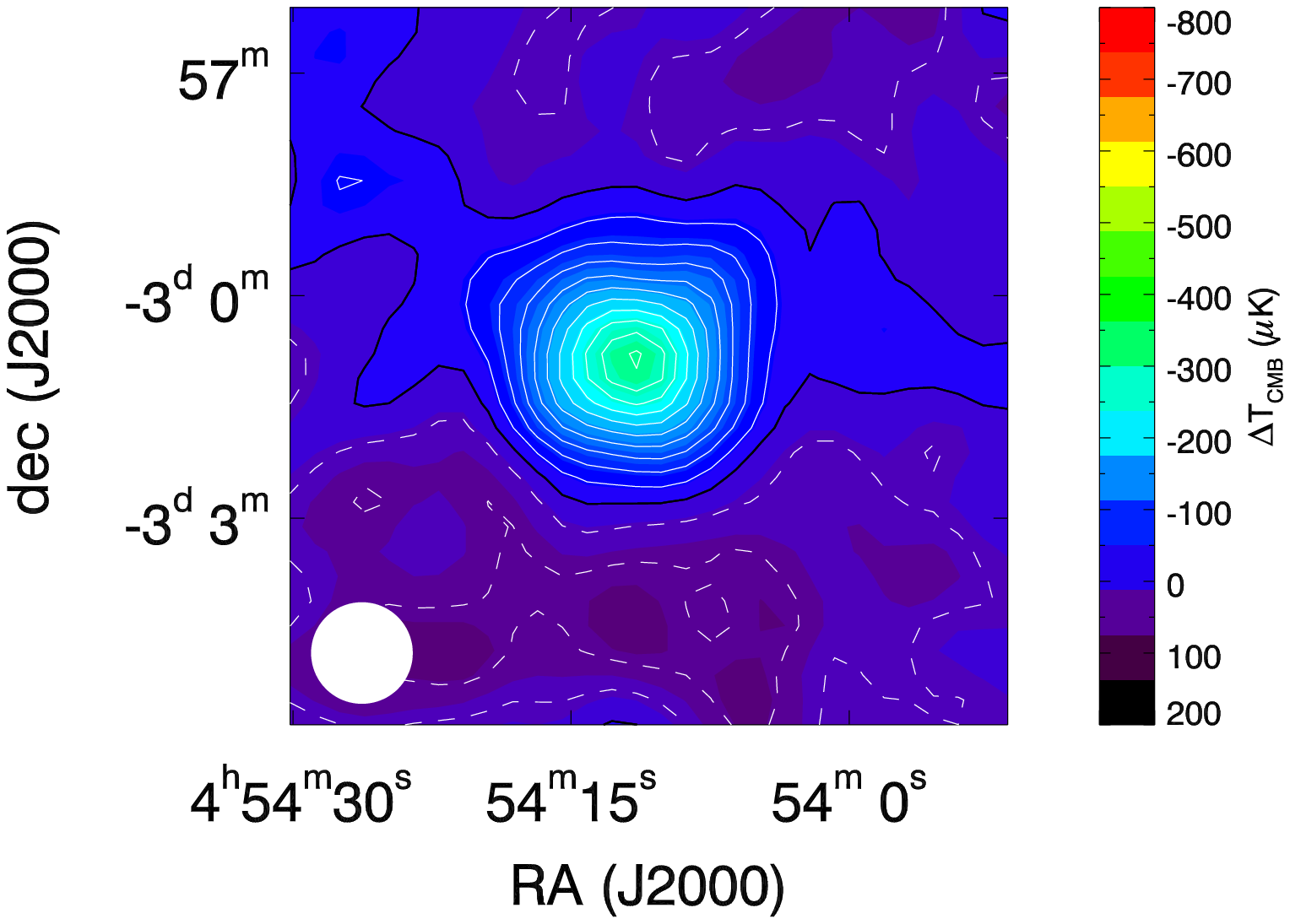} \hspace{.025\textwidth}

    \vspace{.02\textheight}
    \centering
    \includegraphics[height=.25\textheight]{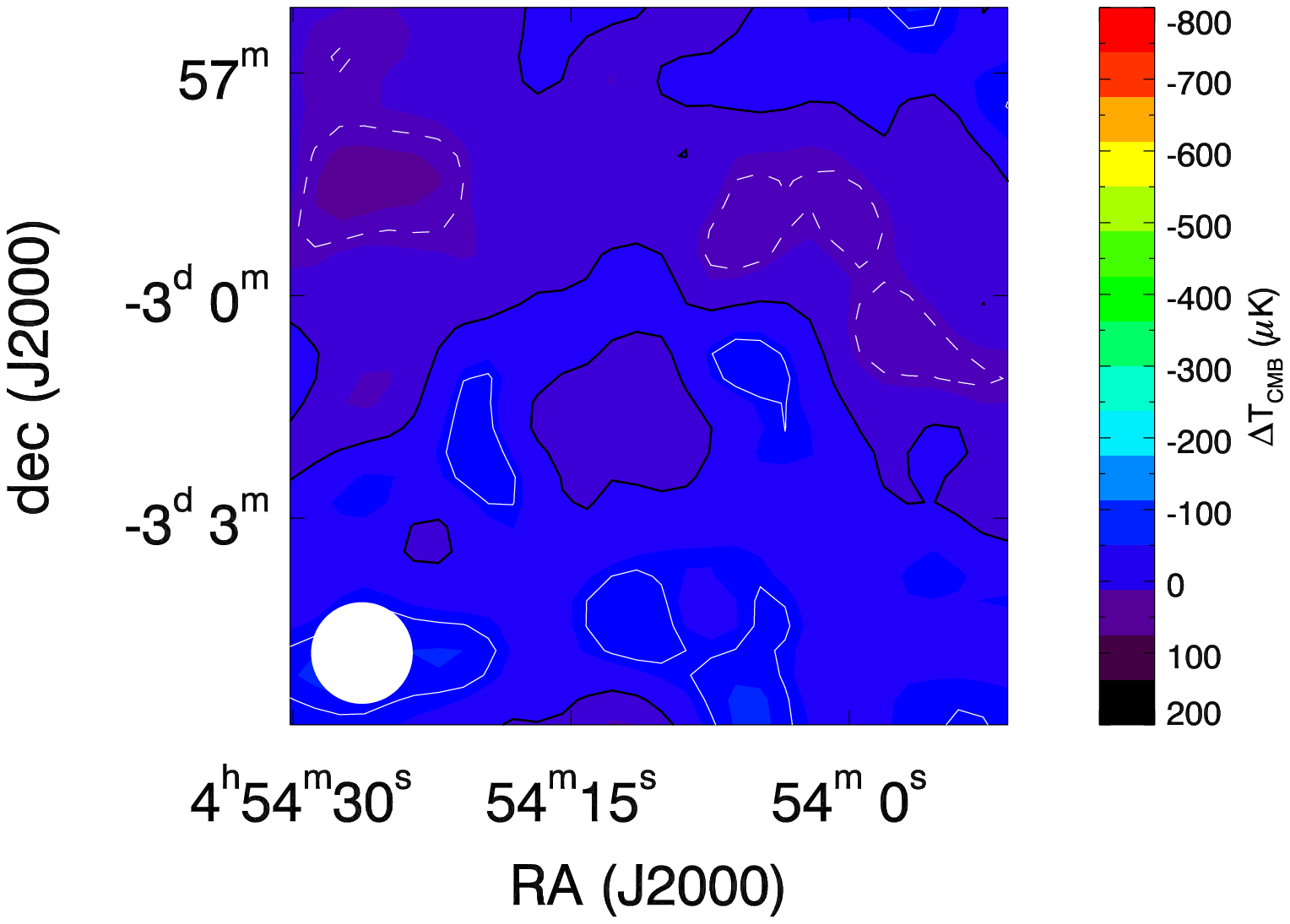} \hspace{.025\textwidth}
    \includegraphics[height=.25\textheight]{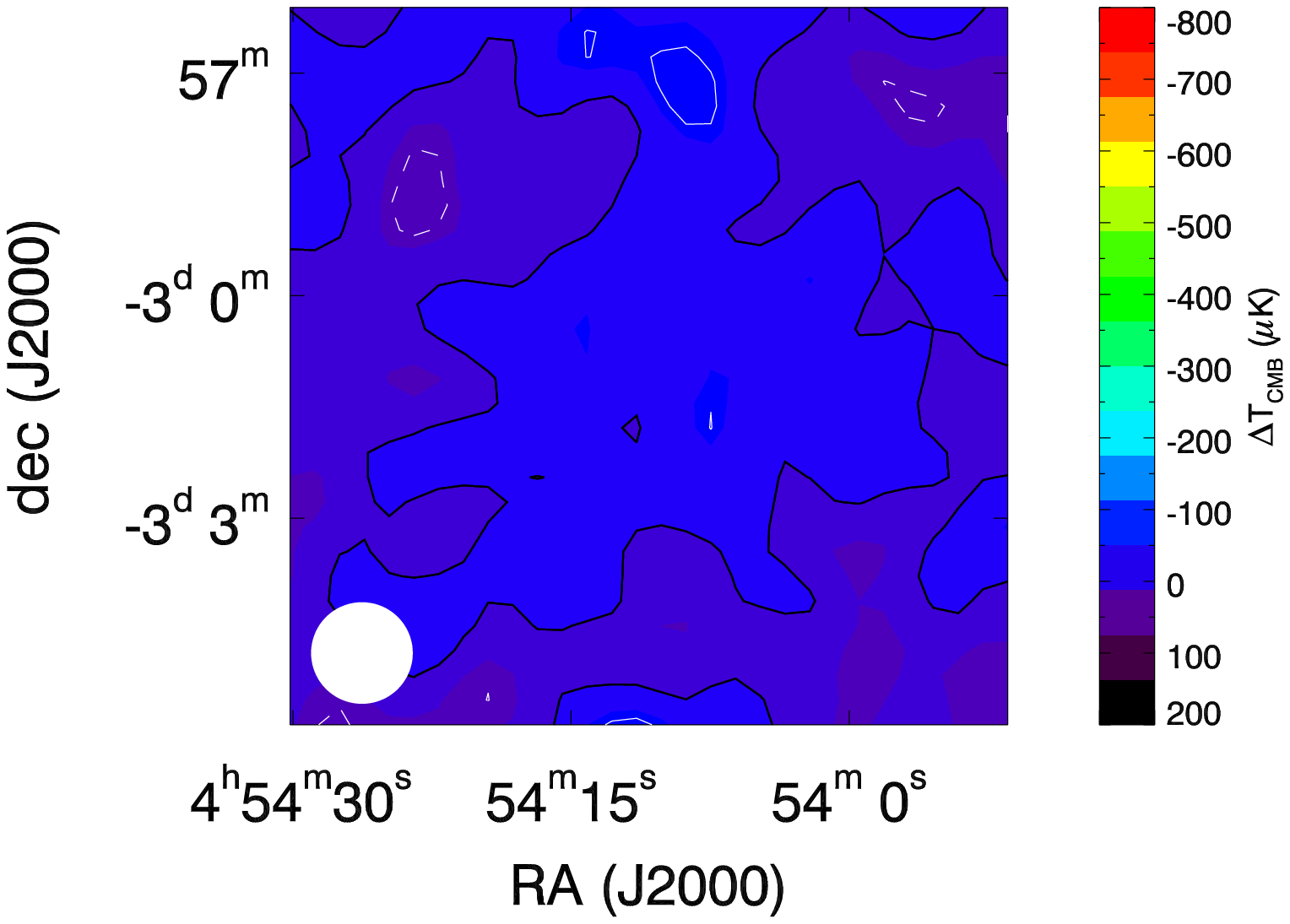} \hspace{.025\textwidth}

    \vspace{.02\textheight}
    \centering
    \includegraphics[height=.35\textheight]{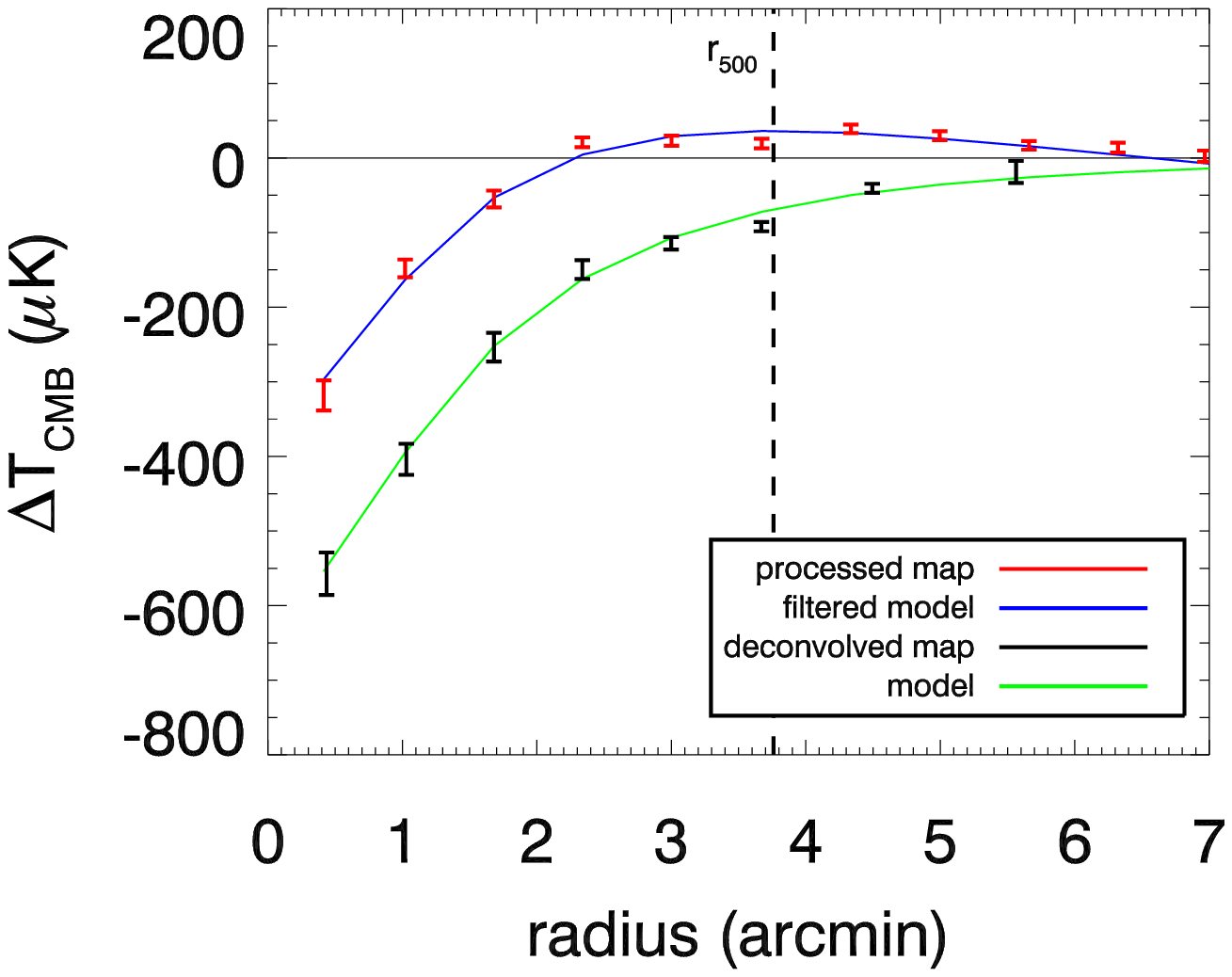}
    \caption{MS 0451.6-0305; from left to right and top to bottom we show
      the deconvovled image of the cluster, 
      the processed image of the cluster,
      the residual map between the processed image of the 
      cluster and the best-fit elliptical Nagai model, 
      one of the 1000 noise realizations for the processed data,
      and a binned radial profile.
      The contour lines represent a S/N of $2,4,..$.}
    \label{fig:ms0451}
  \end{figure*}

  \clearpage
  \begin{figure*}
    \centering
    \includegraphics[height=.25\textheight]{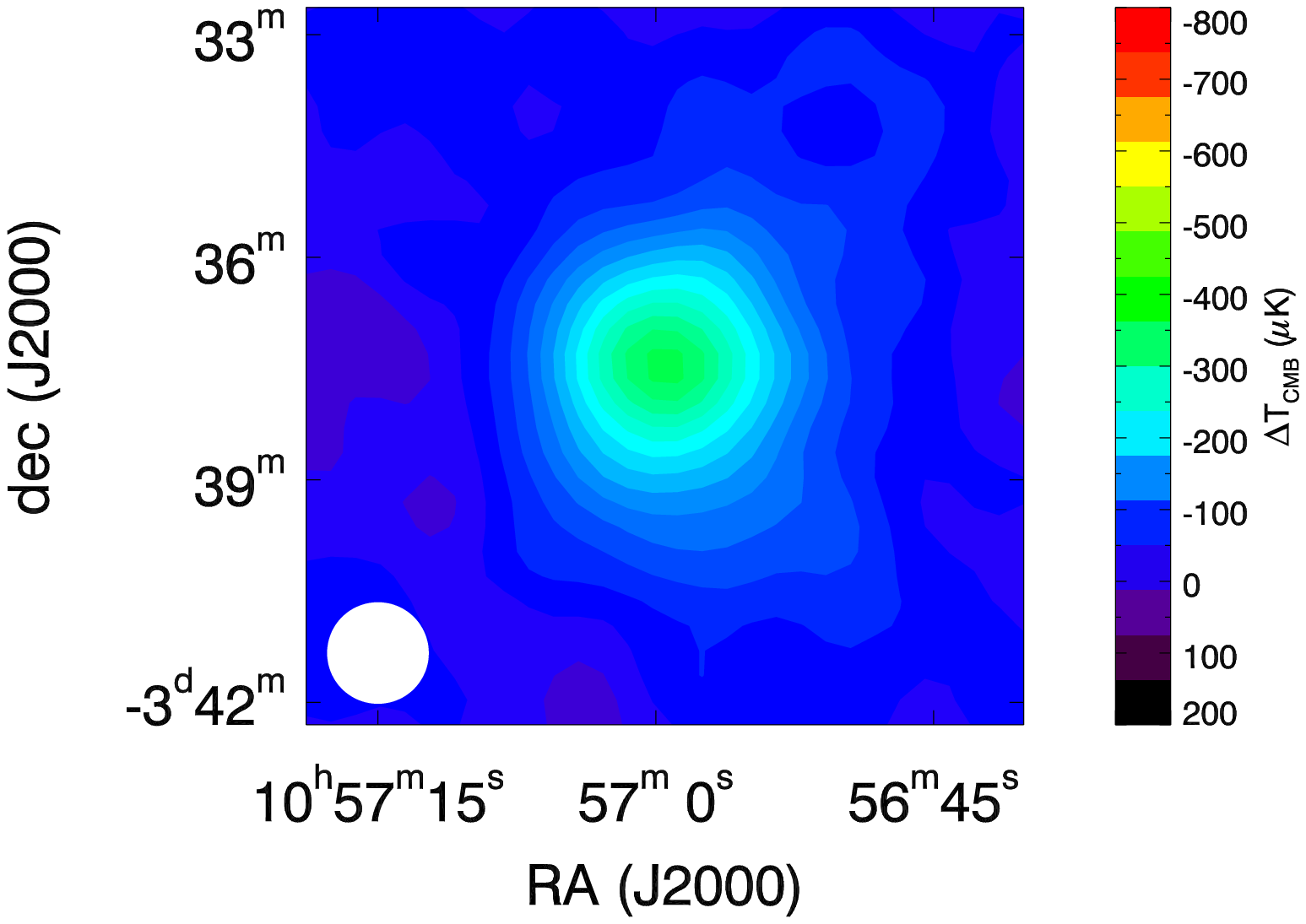} \hspace{.025\textwidth}
    \includegraphics[height=.25\textheight]{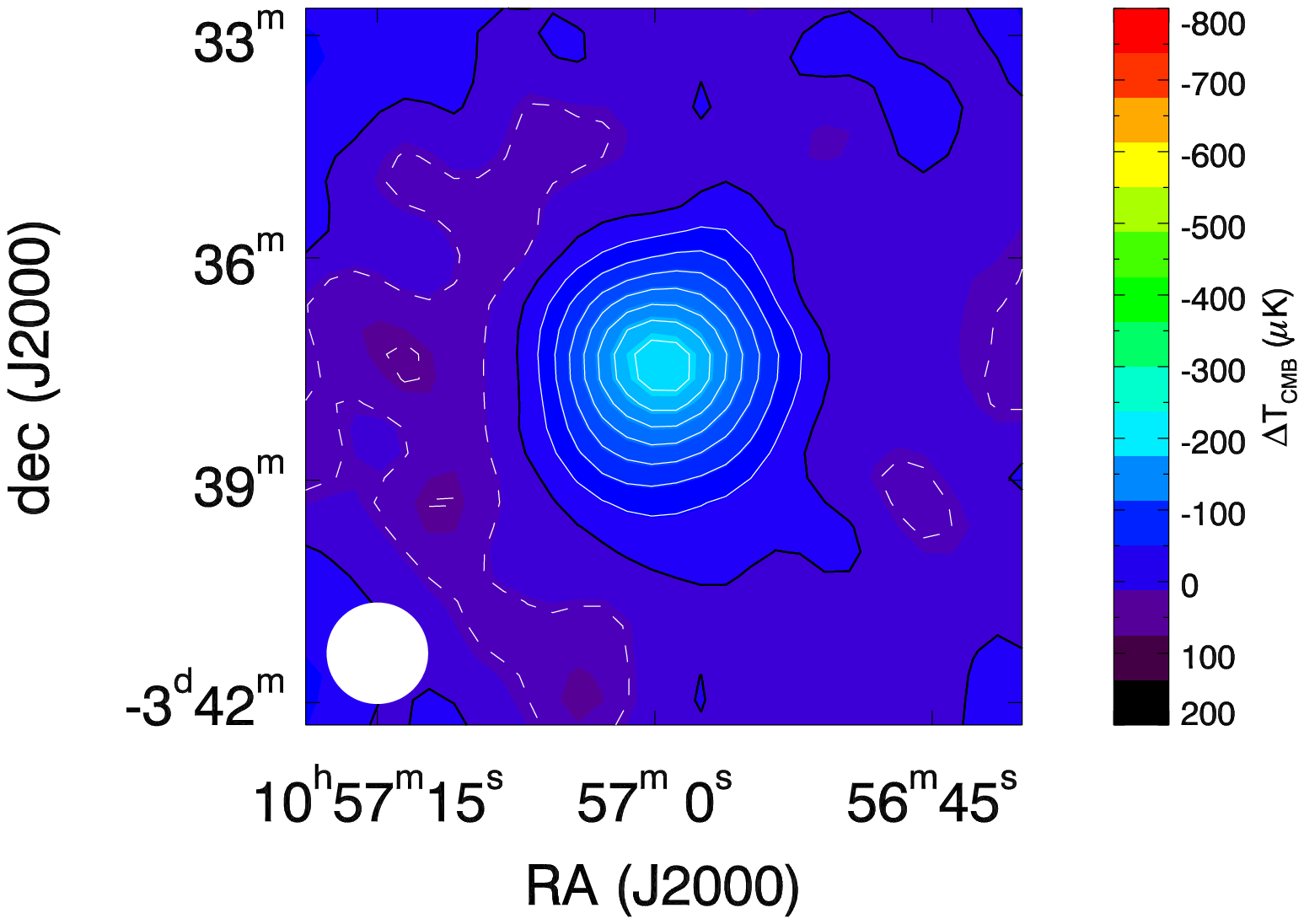} \hspace{.025\textwidth}

    \vspace{.02\textheight}
    \centering
    \includegraphics[height=.25\textheight]{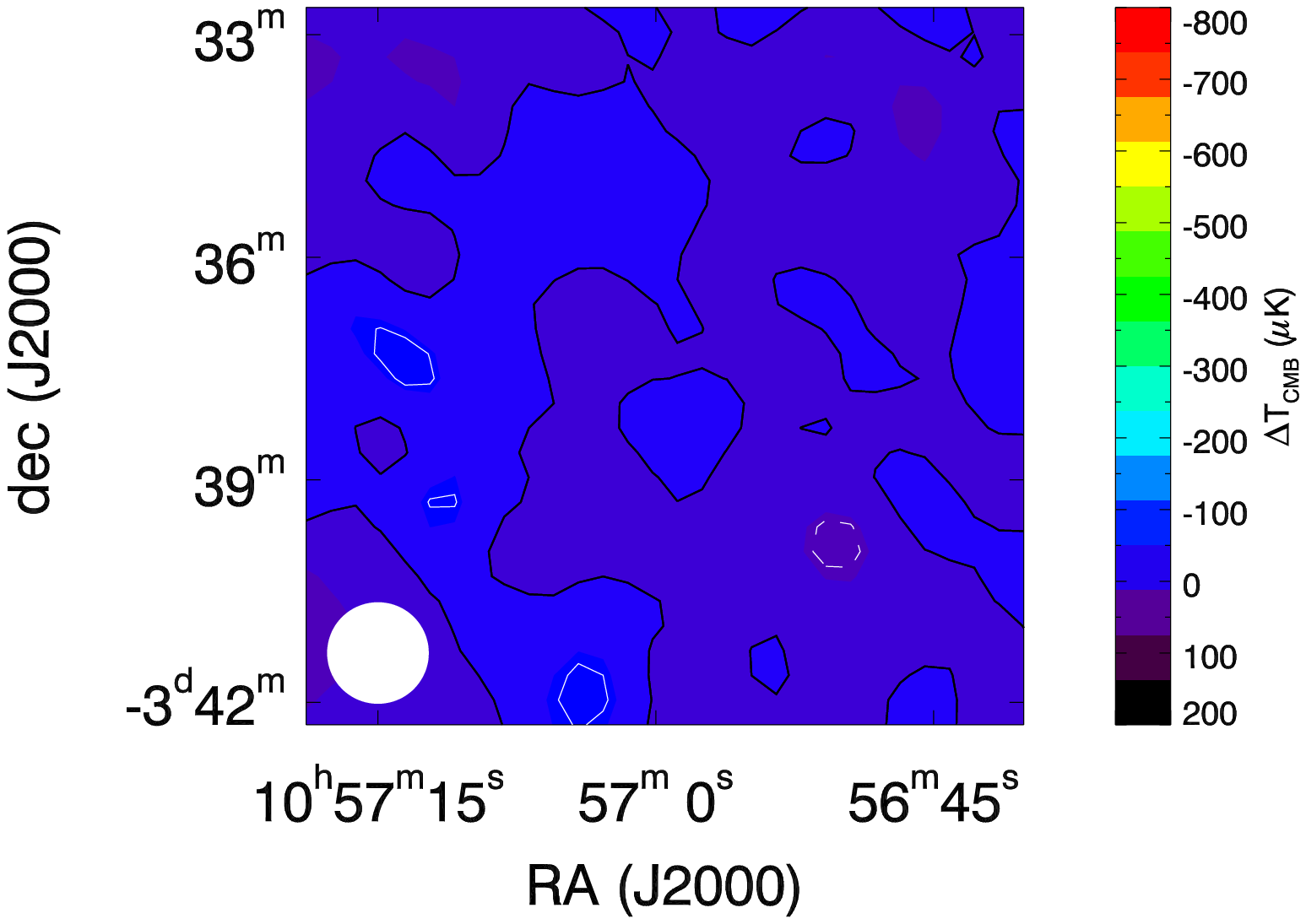} \hspace{.025\textwidth}
    \includegraphics[height=.25\textheight]{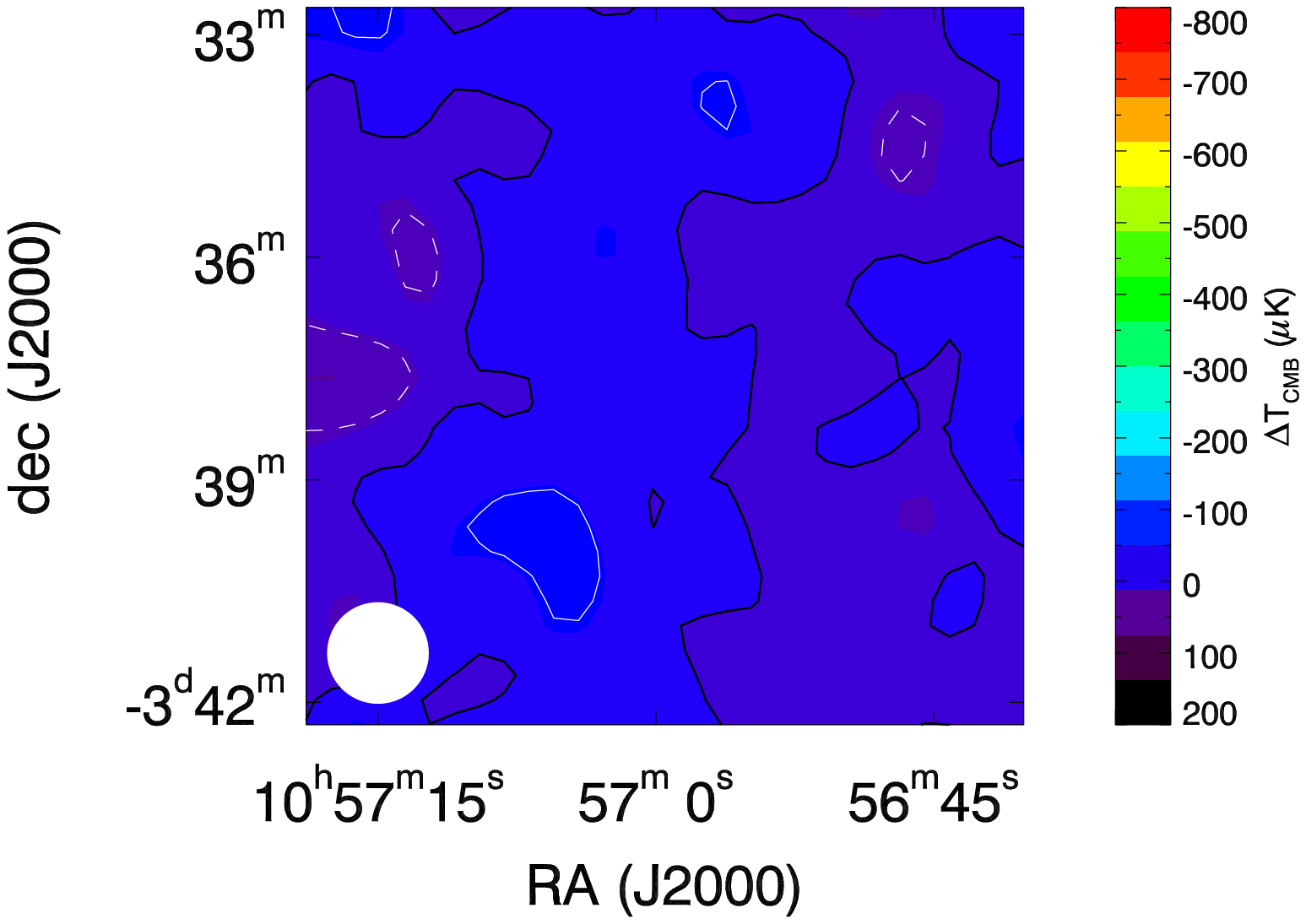} \hspace{.025\textwidth}

    \vspace{.02\textheight}
    \centering
    \includegraphics[height=.4\textheight]{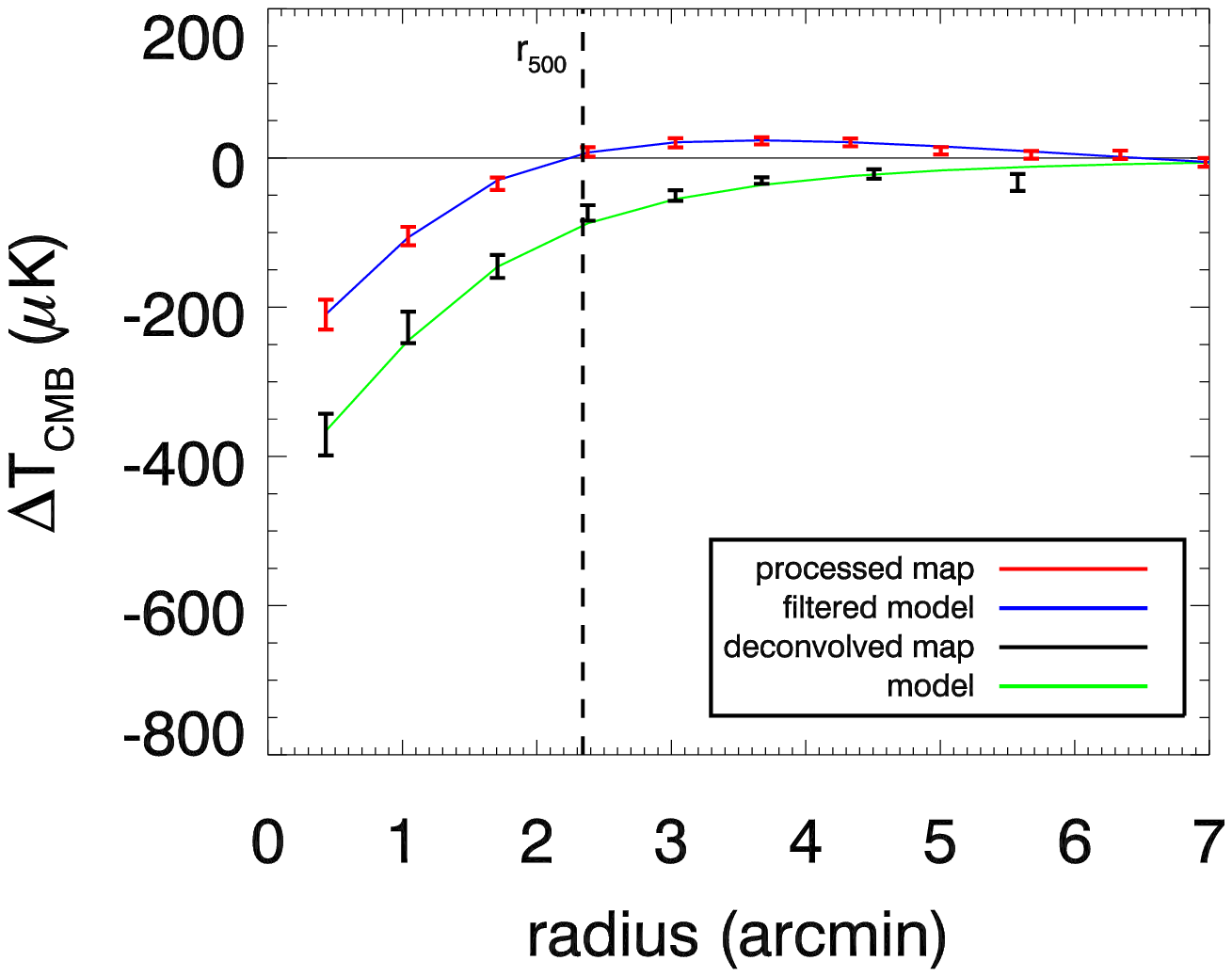}
    \caption{MS 1054.4-0321; from left to right and top to bottom we show
      the deconvovled image of the cluster, 
      the processed image of the cluster,
      the residual map between the processed image of the 
      cluster and the best-fit elliptical Nagai model, 
      one of the 1000 noise realizations for the processed data,
      and a binned radial profile.
      The contour lines represent a S/N of $2,4,..$.}
    \label{fig:ms1054}
  \end{figure*}

  \clearpage
  \begin{figure*}
    \centering
    \includegraphics[height=.25\textheight]{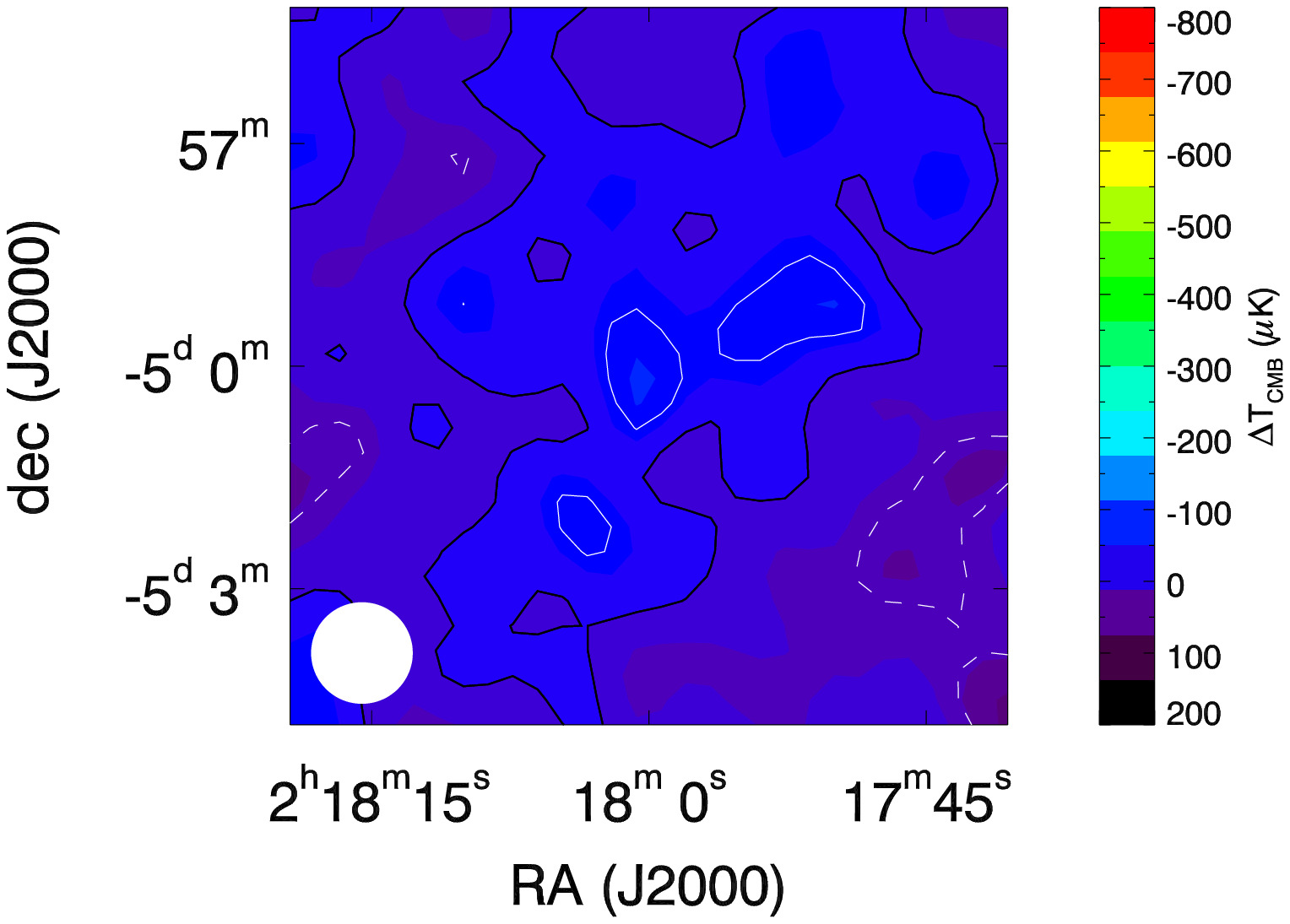} \hspace{.025\textwidth}
    \includegraphics[height=.25\textheight]{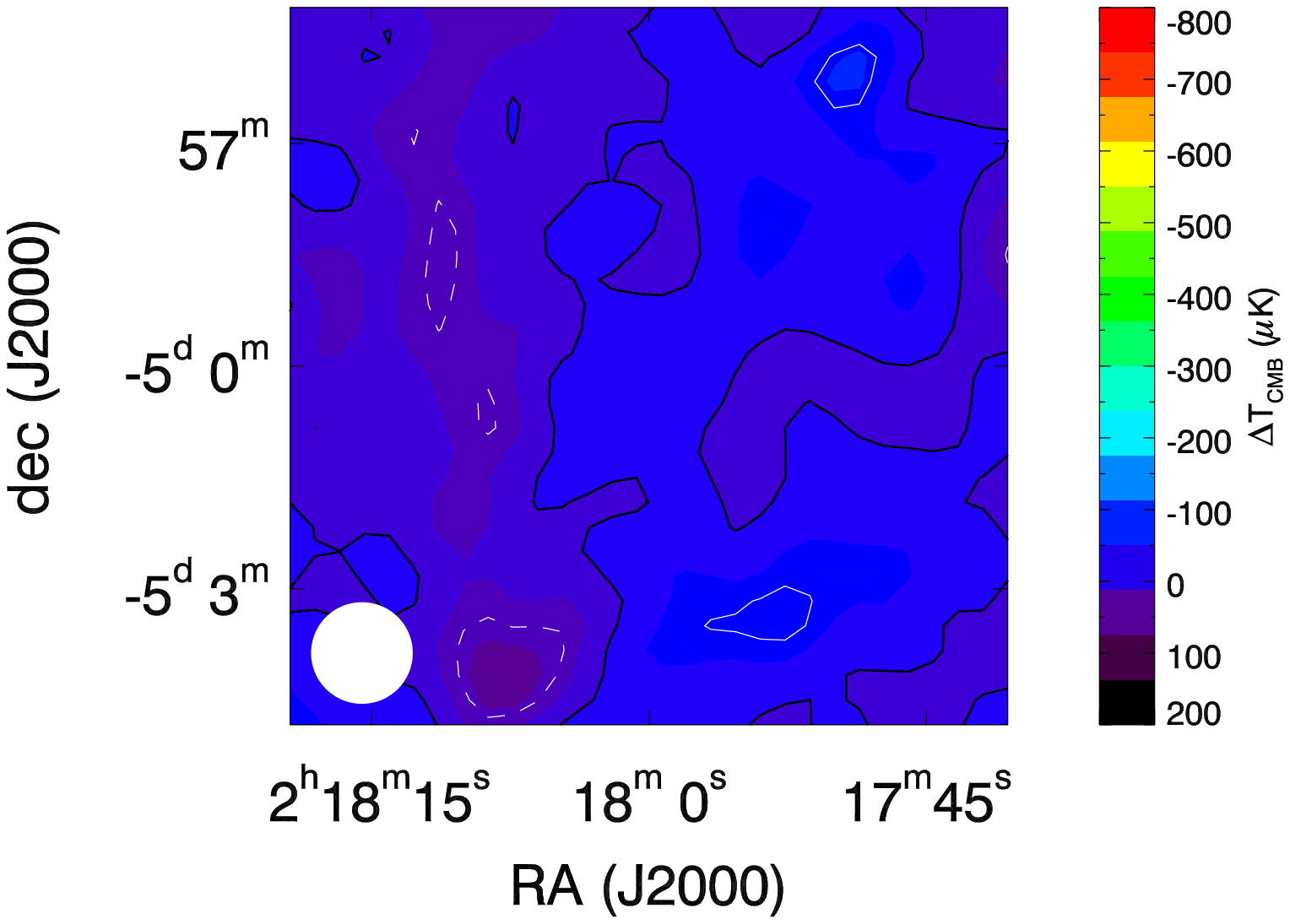} \hspace{.025\textwidth}

    \vspace{.02\textheight}
    \centering
    \includegraphics[height=.4\textheight]{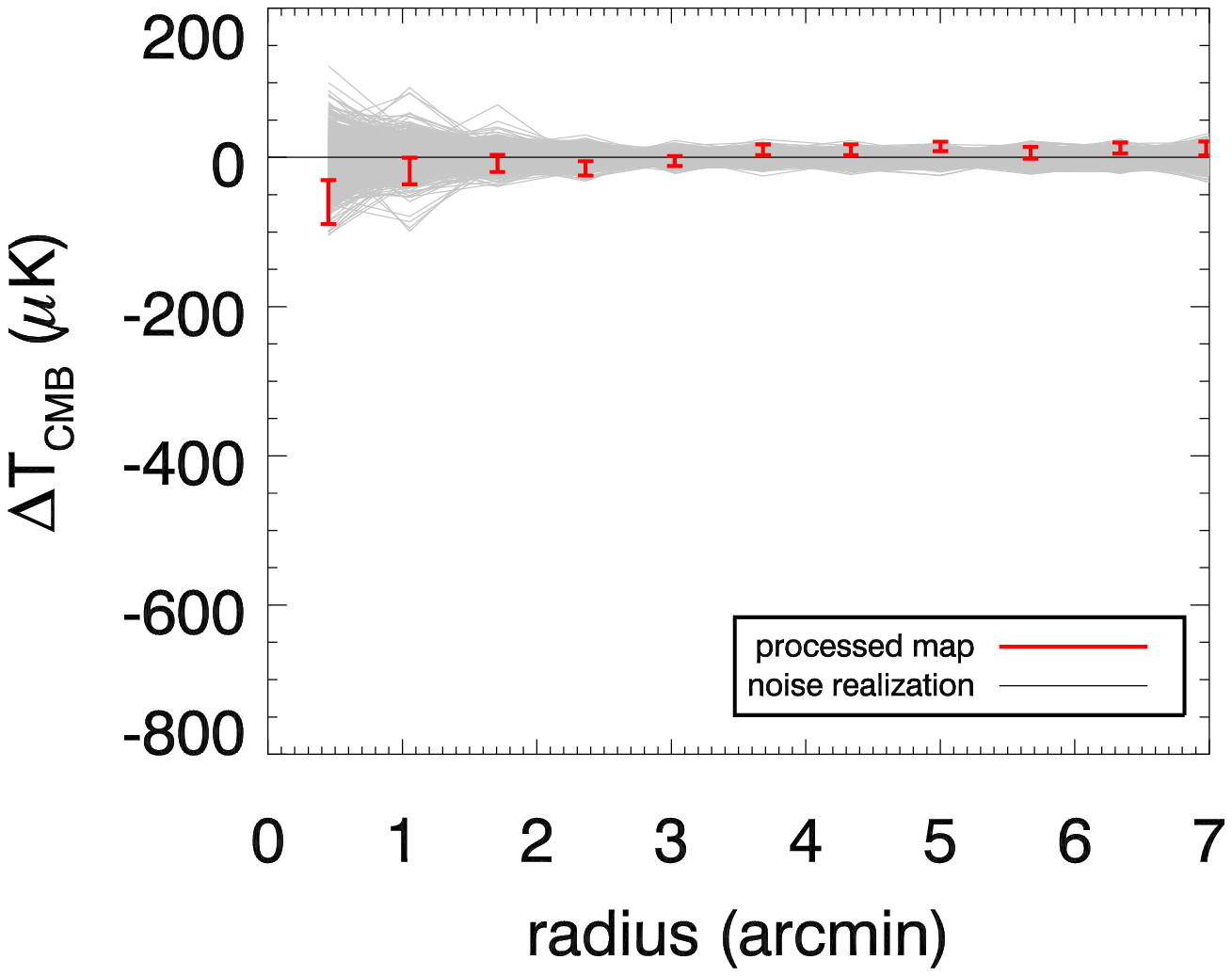}
    \caption{SDS1; from left to right and top to bottom we show
      the processed image of the field, 
      one of the 1000 noise realizations for the processed data,
      and a binned radial profile.
      The contour lines represent a S/N of $2,4,..$.
      The thin grey lines show the radial profiles for each
      of the 1000 noise realizations.}
    \label{fig:sds1}
  \end{figure*}

\end{document}